\documentclass[11pt]{article}
\usepackage[utf8]{inputenc}
\usepackage[titletoc,toc,title]{appendix}
\usepackage{multirow}
\usepackage{tabulary}
\usepackage{booktabs}
\usepackage{authblk}
\usepackage{amsmath}
\usepackage{graphicx}
\usepackage{hyperref}
\usepackage{caption}
\usepackage{subcaption}
\usepackage[normalem]{ulem}
\usepackage{longtable}
\usepackage{afterpage}
\usepackage{bm}
\usepackage{lscape}
\useunder{\uline}{\ul}{}
\usepackage[table,xcdraw]{xcolor}
\usepackage{xcolor}
\usepackage{makecell}
\usepackage{multirow}
\usepackage{rotating}
\usepackage{dsfont}
\usepackage{tikz}
\usepackage{enumitem}
\usepackage{amsfonts}
\usepackage{tabularx}
\usepackage[ruled,linesnumbered]{algorithm2e}
\RestyleAlgo{ruled}
\setlist{nosep}
\usepackage[
backend=biber, natbib,
citestyle = authoryear,
style = authoryear,
uniquelist=false,
uniquename = false,
sorting = nyt,
maxbibnames=99,
maxcitenames=2,
dashed=false
]{biblatex}
\renewbibmacro{in:}{}
\DeclareMathOperator*{\esssup}{ess\,sup}

\newcolumntype{R}{>{\raggedleft\arraybackslash}X}

\title{\bf Incident-Specific Cyber Insurance}

\author[$\dagger$]{Wing Fung Chong}
\author[$\ddagger$]{Dani{\"e}l Linders}
\author[$\star$]{Zhiyu Quan}
\author[,$\star$,$\diamond$]{Linfeng Zhang\thanks{Corresponding author.}}
\affil[$\dagger$]{Maxwell Institute for Mathematical Sciences and Department of Actuarial Mathematics and Statistics, Heriot-Watt University. Email: alfred.chong@hw.ac.uk.}
\affil[$\ddagger$]{Faculty of Economics and Business, University of Amsterdam. Email: d.h.linders@uva.nl.}
\affil[$\star$]{Department of Mathematics, University of Illinois at Urbana-Champaign. Email: \{zquan, lzhang18\}@illinois.edu.}
\affil[$\diamond$]{School of Actuarial Science \& Risk Management, Drake University. Email: linfeng.zhang@drake.edu}
\date{}
\begin{document}
\sloppy

\maketitle

\begin{abstract}
In the current market practice, many cyber insurance products offer a coverage bundle for losses arising from various types of incidents, such as data breaches and ransomware attacks, and the coverage for each incident type comes with a separate limit and deductible. Although this gives prospective cyber insurance buyers more flexibility in customizing the coverage and better manages the risk exposures of sellers, it complicates the decision-making process in determining the optimal amount of risks to retain and transfer for both parties. This paper aims to build an economic foundation for these incident-specific cyber insurance products with a focus on how incident-specific indemnities should be designed for achieving Pareto optimality for both the insurance seller and buyer. Real data on cyber incidents is used to illustrate the feasibility of this approach. Several implementation improvement methods for practicality are also discussed.

\vspace{3mm}
{\bf Keywords:}  Risk management, Cyber insurance, Incident specificity, Statistical learning, Pareto optimality.
%, and corresponding mitigation methods are provided.  
% In the current market practice, most of cyber insurance policies solely provide coverages for particular types of cyber losses, without any scientific foundation.This paper explores the possibility of introducing incident-specific cyber insurance policies. 
% This paper 
% In order to do so, a thorough cyber risk assessment is necessary, in terms of the multi-class classification problem for cyber incident type given incident characteristics, and conditional severity modeling. This paper provides a data-driven justification for providing such incident-specific cyber insurance contracts for the market substantiality and potential growth.
\end{abstract}

\section{Introduction}
In May 2021, Colonial Pipeline, one of the largest oil pipeline systems in the United States (US), was hit by a ransomware attack and had a shutdown for five days; see \citet{OfficeCybersecurity2021}. In November 2020, Amazon Web Services, a major cloud service provider that many businesses rely on, had a severe outage triggered by an operating system configuration; see \citet{Greene2020}. In 2017, Equifax, one of the three largest US credit reporting agencies, experienced a major data breach that exposed the private records of nearly 150 million American citizens; see \citet{Equifax2017}. In 2016, a class action was brought against Meta Platforms (formerly known as Facebook) for failing to comply with the Biometric Information Privacy Act in Illinois, and the case was settled for \$650 million in 2021; see \citet{Holland2021}. These examples illustrate the extraordinary exposure of modern businesses to cyber risk and their possible devastating consequences. 

The events listed above represent a wide range of incidents that are considered cyber-related, and they are common in this digital world where business operations rely heavily on data and cyber systems. To mitigate the aftermath impact of such incidents, cyber insurance products are designed for businesses to purchase. The insurability of cyber losses is examined in \citet{dacorogna_building_2023}, in which their loss expectations are shown to be finite. These cyber insurance products typically cover losses resulting from various perils. As summarized in \citet{Romanosky2019, Marotta2017, Woods2017}, some cyber insurance policies indemnify losses caused by ransomware, and some policies cover liabilities resulting from data breaches. It is also not unusual for a single policy to cover multiple types of cyber incidents. In this paper, our focus is on those policies that offer coverage for multiple perils, which are referred to as \textit{incident-specific cyber insurance}.

\subsection*{Need for incident-specific cyber insurance}
Incident-specific cyber insurance products are in line with businesses' needs for cyber risk management. 

A definition of cyber risk commonly adopted in practice and literature is ``cyber risks are operational risks to information and technology assets that have consequences affecting the confidentiality, availability, and integrity of information and information systems''; see \citet{Cebula2010}. Aligning to this idea, the National Institute of Standards and Technology (NIST) Cybersecurity Framework, from which many cybersecurity standards for critical infrastructure sectors are derived, considers data security from the perspectives of data confidentiality, integrity, and availability and proposes a comprehensive set of best practices; see \citet{NIST2018}. In \citet{Eling2019}, the authors argued that cyber risks are subcategories of operational risk and used cyber-related data from a database of operational losses to model the loss of four types of cyber incidents, including the ones caused by actions of people, system failure, internal process failure, and external events. In \citet{Amin2017}, the author considered the same four categories of cyber risks and used Bayesian networks to model the relationship between risk categories and risk factors. In \citet{Ghadge2019}, the authors proposed a different set of incident types for supply chain cyber risk management, including physical threats, breakdown, indirect attacks, direct attacks, and insider threats. 

Provided that cyber risk is a collective term referring to the risks of multiple types of cyber incidents, and businesses typically demand a risk management strategy that is holistic and, at the same time, can address the subtle differences among various incident types, purchasing an incident-specific cyber insurance policy is then a natural choice for businesses to meet such a need.

\subsection*{Existing incident-specific cyber insurance}
To ensure risk exposures are manageable, many incident-specific cyber insurance policies come with sublimits and separate deductibles for individual coverages. For example, \citet{AIG2019} allows for setting limits and deductibles individually for events such as cyber extortion and computer crime. \citet{Romanosky2019} surveyed 67 policies and found many of them use sublimits. In this paper, these sublimits and deductibles shall be referred to as \textit{incident-specific limits}, or simply \textit{limits}, and \textit{incident-specific deductibles} or \textit{deductibles}. The scope of this study though does not include the cases in which there is a limit or deductible at the aggregate policy level, which shall be deferred to future research.

Although insurers can shield them from excessive exposures using those risk-sharing provisions, such policy designs complicate the process of shopping for cyber insurance coverage for prospective buyers. Given that most companies have difficulties in determining the appropriate amount of coverage (see \citet{johansmeyer_cybersecurity_2021}), this problem will only be more of an obstacle that hinders cyber insurance purchases if companies have to impromptu configure the amounts of coverage for individual incident types.  

\subsection*{Proposed workflow for determining coverage}
Provided that incident-specific cyber insurance is a norm in practice but lacks a technical foundation that justifies the determination of coverage amounts,
it is in the interests of both insurers and companies, who seek insurance coverage, to make the process of determining incident-specific coverage easy to understand, compatible with the existing underwriting procedures, and mutually beneficial to both parties. To this end, we shall address the aforementioned problem of determining incident-specific coverage from a bilateral perspective. 

Drawing inspiration from \citet{Asimit2021}, which shows that an environment-specific indemnity profile could be Pareto optimal for both the buyer and seller, in this paper, we propose an economically sound workflow that determines the appropriate amount of incident-specific coverage. By doing so, it helps the insured and the insurer reach a Pareto optimality, such that the risk taken by either of the two parties cannot be further reduced without increasing the other's risk. How such a problem should be formulated and how both parties' risks are measured are elaborated in Section \ref{sec:formulation}.

\begin{figure}[htbp]
    \centering
    \includegraphics[width=\textwidth]{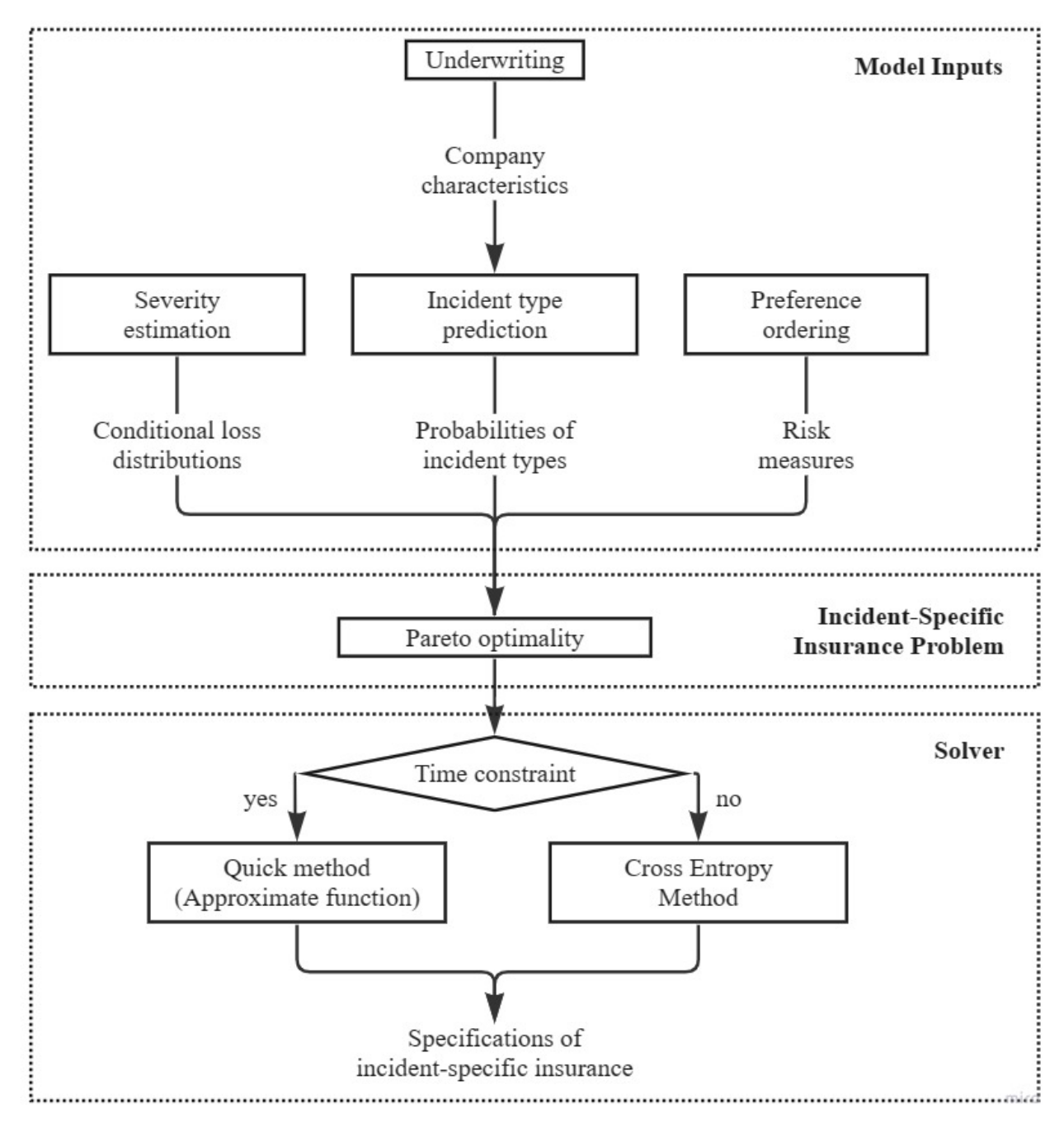}
    \caption{Workflow of designing incident-specific cyber insurance coverage}
    \label{fig:workflow_diagram}
\end{figure}

The diagram in Figure \ref{fig:workflow_diagram} illustrates this workflow, which consists of three major parts: (I) the \textit{incident-specific cyber insurance problem}, which serves as the central pillar of this workflow and assigns optimal amounts of risk to the policyholder and the insurer; (II) \textit{model inputs}, in which severity and incident type can be estimated and statistically learned based on historical cyber incident data and the underwriting process, together with the policyholder's and insurer's preferences information for the decision-making part; and (III) the \textit{solver} for the optimal insurance problem, which can be implemented using either an exact method or an approximated but quick method depending on whether there is a time constraint. The solver produces specifications of the incident-specific insurance coverage, which is the final and desired outcome of this workflow.
%Each of these three parts shall be elaborated in this paper. 

Regarding Part (III), a practical scenario is that the insurer needs to provide a quote and a coverage to a prospective customer on-the-fly. We shall show that using the exact method may result in a long waiting time, and thus the quick method, which generates the results almost instantly, can be a workaround in that situation.

In this paper, we shall detail the key components of the proposed workflow for designing incident-specific cyber insurance coverage, and then demonstrate it with numerical examples.

\subsection*{Findings and contributions}
First, based on the incident-specific insurance design problem, the cyber insurance contract supplied by the proposed workflow must be reducing the aggregate cyber risk of a company, as an insurance buyer, and the insurer after the risk transfer; the contract must also be mutually beneficial to both the insured and the insurer since a paid premium is less than the buyer's cyber risk reduction while is able to cover the risk taken by the insurer. However, their aggregate cyber risk reduction is not necessarily effective, in the sense that some insurance buyers might pay more premium than their reduction in the aggregate risk with the insurer; this is indeed observed from the numerical examples presented in Section \ref{sec:fun_approx}. 
% from the numerical examples, for some cyber insurance buyers, their paid premium would always be more than the reduction in the aggregate risk.
%First, while, based on the incident-specific insurance design problem, a company, as an insurance buyer, experiences a risk reduction greater than a paid premium which is able to cover the taken risk by the insurer, for some companies, an appropriate amount of premium for the incident-specific insurance contract supplied by the proposed workflow could be less than the reduction of aggregate risk taken by both parties after the risk transfer. This illustrates the effectiveness of the incident-specific cyber insurance described in this paper in reducing cyber risk and being mutually beneficial to both the insured and the insurer.
% {\color{red}One is that when a company, as an insurance buyer, pays an appropriate amount of premium for an incident-specific insurance contract supplied by the proposed workflow, the company will experience a risk reduction greater than the paid premium. On the other hand, although the insurer has an increased risk for covering the insured's losses, the collected premium exceeds the taken risk. Consequentially, the aggregate risk taken by both parties is reduced. }
Second, the proposed workflow is computationally tractable in scenarios that require the specifications of an incident-specific policy to be delivered promptly, such as quick quote and coverage services, comparing to other conventional methods.

Therefore, the contributions of this paper are as follows. Firstly, cyber insurance policies commonly cover multiple perils, but the determination of incident-specific coverage amounts lacks justification in practice and literature. The proposed workflow leads to insurance specifications that are justified and economically sound to both the insured and the insurer. Secondly, the proposed workflow is data-driven and compatible with the existing underwriting procedures. All the data required by this workflow can be obtained from public or proprietary sources by underwriters, which results in no additional work on the frontend. Thirdly, we show that, from the numerical examples, while the designed incident-specific cyber insurance contract must be mutually beneficial to the insured and the insurer, their aggregate cyber risk reduction might not be always effective. 
%{\color{red}We also showed using real data that an incident-specific cyber policy as an outcome of the proposed workflow makes both the insured and the insurer better off; \textit{i.e.}, the insured enjoys a reduced risk greater than the paid premium, and the insurer collects more premium than the risk taken.}
Lastly, this paper overcomes the possible computational challenges arising from the implementation of such a workflow, making it versatile and adaptable to different scenarios depending on the amount of time permitted to produce results. 

The rest of this paper is organized as follows. Section \ref{sec:formulation} formulates the problem of designing incident-specific cyber insurance as an optimization problem. Section \ref{sec:model_inputs} discusses the method and the data that we used to obtain the necessary model inputs. Given the model inputs, Section \ref{sec:cme} describes the Cross Entropy Method and how it is used to solve the proposed problem as well as presents and discusses some numerical results. Section \ref{sec:fun_approx} presents the use of function approximation to address some computational challenges in the process of solving the optimization problem. We conclude in the last section with discussions of potential applications and future works.

% A survey conducted by PartnerRe (2018) shows that one of the major complaints that policyholders have about cyber insurance is the insufficient coverage. The insufficiency is two-fold. One is that many cyber insurance policies have a limited scope of coverage in terms of the covered loss types. The other is that there is a capacity constraint in the market, which may prevent policyholders from getting enough coverage from a single insurer. 

\section{Design of Incident-Specific Cyber Insurance} \label{sec:formulation}

\subsection{Buyer and seller's Pareto optimality}
Incident-specificity in this study means that although the insurance product is designed as a package with coverage for a variety of cyber incident types, the coverage for each incident type has its own indemnity function. In addition, a major characteristic of the policy design is that different types of cyber incidents are mutually exclusive, \textit{i.e.}, every incident precisely belongs to a unique category. Therefore, the occurrence of one incident would only trigger one and only one corresponding incident-specific coverage. This point will be further illustrated with real data in later sections. Then, the problem of interest in this study is how to determine the incident-specific indemnity functions. 

To formulate this problem, let $(\Omega,\mathbb{F})$ be a measurable space, let $\mathbb{P}$ be a probability measure on $(\Omega,\mathbb{F})$, and let $\mathcal{K}=\left\{1,2,\dots,K\right\}$ be the set of indices of insurable incident types with cardinality $K$. For a cyber incident of any type, denote the random variable of the realized non-negative loss as $(X)\mathbb{I}_{\{O=k\}}$, where $X$ is the loss resulting from a particular cyber event and the indicator function $\mathbb{I}_{\{O=k\}}$ takes the value of $1$ if the \textbf{O}ccurred incident type $O$ is $k$, for any $k \in \mathcal{K}$, and takes the value of $0$ otherwise. Furthermore, let $I_k$ be the incident-specific \textbf{I}ndemnity function, for incident type $k\in\mathcal{K}$; that is, the insurance buyer receives $I_k(X)\mathbb{I}_{\{O=k\}}$ after an incident. The amount of loss \textbf{R}etained by the buyer is $R_k(X)\mathbb{I}_{\{O=k\}} = (X)\mathbb{I}_{\{O=k\}} - I_k(X)\mathbb{I}_{\{O=k\}}$, which could be non-zero in the scenario that an incident-specific indemnity does not fully cover the corresponding loss.

Acquiring such an incident-specific insurance policy costs the buyer a single premium $\pi$. From the buyer's perspective, the total loss at the end of the policy period is, $\pi+\sum_{k=1}^K R_k(X)\mathbb{I}_{\{O=k\}}$, where the summation is a result of the mutual exclusivity of incident types. The risk-sharing counterparty, \textit{i.e.}, the seller, correspondingly bears the risk of value $\sum_{k=1}^K I_k(X)\mathbb{I}_{\{O=k\}}$ in exchange for receiving the premium $\pi$, thus making the loss at the end of the policy period, $-\pi+\sum_{k=1}^K I_k(X)\mathbb{I}_{\{O=k\}}$. In this study, any potential return generated by the premium during the policy period is neglected for simplicity.
%For simplicity, the potential return generated by the premium paid at the beginning of the period is neglected, and it shall be shown that it does not affect the decision on contract design.   

To support the decision-making regarding configurations of the contract consisting of the incident-specific indemnity functions and the policy premium, assume that the \textbf{B}uyer and the \textbf{S}eller are endowed with risk measures $\rho_B$ and $\rho_S$, respectively, and the {\it Pareto optimality} is to be achieved with respect to the risk measures on their post-transfer risk positions, so that neither of the two parties can attain a lower risk, evaluated by the risk measure, without the other party's risk being increased. Then, as discussed in \citet{Asimit2021} and proved in Theorem 3.1 of \citet{Asimit2018}, if their risk measures are translation invariant (which are satisfied by most risk measures), their Pareto optimality is achieved by the contract which minimizes:
\begin{equation*}
    F(I_1,I_2, \dots, I_K,\pi; X, O) = \rho_B\left(\sum_{k=1}^K R_k(X)\mathbb{I}_{\{O=k\}}+ \pi \right) + \rho_S\left(\sum_{k=1}^K I_k(X)\mathbb{I}_{\{O=k\}} -\pi\right).
\end{equation*}

With their risk measures being translation invariant and the policy premium $\pi$ being a deterministic variable, the objective function, with a slight abuse of notation, can be simplified to
\begin{equation*}
    F(I_1,I_2, \dots, I_K; X, O) = \rho_B\left(\sum_{k=1}^K R_k(X)\mathbb{I}_{\{O=k\}} \right) + \rho_S\left(\sum_{k=1}^K I_k(X)\mathbb{I}_{\{O=k\}} \right),
\end{equation*}
The minimization problem that solves the ex-ante specified Pareto optimal incident-specific indemnity functions is then given by:
\begin{equation} \label{eq:obj}
    \min_{(I_1,I_2, \dots, I_K) \in \mathcal{I}} F(I_1,I_2, \dots, I_K; X, O),
\end{equation}
where, with $\mathrm{Id}$ being the identity function, the solution space 
\begin{equation*}
\mathcal{I}:= \{(I_1,I_2, \dots, I_K) : 0 \le I_k \le \mathrm{Id},\;I_k \text{ and } R_k \text{ are non-decreasing},\;\forall k\in\mathcal{K}\}
\label{eq:mathcalI}
\end{equation*}
adheres to two fundamental principles in insurance, including 1) the buyer is compensated partially or fully for a loss, but cannot make a profit from the compensation; and 2) the ex-post moral hazard that a falsely larger claim is made to reduce the buyer's deductible or the seller's indemnity payout should be avoided.

Finally, the ex-ante specified Pareto optimal policy premium is given by the rationality constraints of the buyer and seller:
\begin{equation*}
\rho_B\left(\sum_{k=1}^K R_k(X)\mathbb{I}_{\{O=k\}}+ \pi \right)\leq\rho_B\left(X\right),
\end{equation*}
\begin{equation*}
\rho_S\left(\sum_{k=1}^K I_k(X)\mathbb{I}_{\{O=k\}}- \pi \right)\leq\rho_S\left(0\right)=0,
\end{equation*}
in which the right-hand-side of the inequalities are the risk measures on both parties' pre-transfer risk positions. By the translational invariance of their risk measures, these can be simplified as:
\begin{equation}\label{eq:premium_range}
\rho_S\left(\sum_{k=1}^K I_k(X)\mathbb{I}_{\{O=k\}}\right)\leq\pi\leq\rho_B\left(X\right)-\rho_B\left(\sum_{k=1}^K R_k(X)\mathbb{I}_{\{O=k\}}\right).
\end{equation}
in which the incident-specific indemnity and retained functions are Pareto optimal being solved in \eqref{eq:obj}. Thus, at the Pareto optimality, the policy premium is not disentangled from the respective risks taken by the buyer and the seller. Varying the Pareto optimal policy premium choice within this interval traces the entire Pareto frontier. Note that, in this paper, the premium is derived from an economic perspective; in the insurance practice, an insurer could charge a premium based on an actuarial price.
% Note that this premium range is derived in an economically feasible manner, and it does not necessarily satisfy the actuarial rationale. In the insurance practice, actuaries may need to use a different approach for pricing. 

Therefore, in the remainder of this study, the focus is on the design of the incident-specific indemnity functions.

\subsection{Optimal indemnities with Value-at-Risk preferences}
The choice of risk measures, $\rho_B$ and $\rho_S$, depends on the insurance buyer's and seller's own risk management appetites. Some common choices, which satisfy the translational invariance, include Value-at-Risk (VaR) and Tail Value-at-Risk (TVaR).

It can be verified that if both $\rho_B$ and $\rho_S$ are the same type of subadditive risk measures and only differ in their parameters (\textit{e.g.}, TVaR with different risk tolerance level parameters), then the solution to Problem \eqref{eq:obj} assigns all risks to the party that has a higher risk tolerance. VaR, on the other hand, is not subadditive but is still prevalently adapted in the banking and insurance industries; see, for example, the capital requirement in Basel III and Solvency II frameworks. Therefore, in this study, we assume that the buyer and the seller both adopt VaR as their risk measures. VaR is as follows:
\begin{equation} \label{eq:VaR}
    \text{VaR}_\gamma(Y) = \inf\{y \in \mathbb{R}: \mathbb{P}(Y \le y) \ge \gamma\},
\end{equation}
where $Y$ is a random variable and $\gamma \in (0,1)$ is the risk tolerance level.

% In this study, we shall consider the scenario that both the buyer and the seller use VaR in contract negotiation and demonstrate how to determine the optimality in that case. 

With the VaR risk preferences, \citet{Asimit2021} showed that the indemnity functions in the sub-solution space $\mathcal{I}\setminus \mathcal{I}_1$ are at least suboptimal for Problem \eqref{eq:obj}, where
\begin{equation}
\begin{aligned}
    \mathcal{I}_1:= \{(I_1,I_2, \dots, I_K) \in \mathcal{I}:&\;  I_k(X) = (X-d_k)_+ \text{ or } I_k(X)= X-(X-d_k)_+, 
    \\ 
    &\;d_k \in [0, \esssup(X)], \text{ for each } k\in\mathcal{K}    \},
\end{aligned}
\label{eq:mathcalI_1}
\end{equation}
% the solution space \eqref{eq:mathcalI} could be narrowed down to 
where $\esssup(X)$ is the essential supremum of $X$ under the probability measure $\mathbb{P}$. 
%, and all solutions in $\mathcal{I}\setminus \mathcal{I}_1$ are at least suboptimal.
This implies that, for each incident-specific coverage, either a deductible or a policy limit, denoted by $d_k$, should be implemented. Within the solution space in \eqref{eq:mathcalI_1}, solving the infinite-dimensional Problem \eqref{eq:obj} is thus reduced to solve the following finite-dimensional, but combinatorial, minimization problem:
\begin{equation} \label{eq:varobj}
        \min_{\substack{\bm{d} \in \mathbb{R}_+^K, \\ \bm{\theta} \in \{0,1\}^K}} F(\bm{d}, \bm{\theta}; X, O) =  \min_{\substack{\bm{d} \in \mathbb{R}_+^K, \\ \bm{\theta} \in \{0,1\}^K}} \left[\text{VaR}_{\alpha}\left(L_S(\bm{d}, \bm{\theta}; X, O)\right) 
        + \text{VaR}_{\beta}\left(L_B(\bm{d}, \bm{\theta}; X, O)\right)\right],
\end{equation} 
where 
\begin{itemize}
    \item $\bm{\theta} = (\theta_1,\theta_2, \dots, \theta_K)$, and for each $k\in\mathcal{K}$, $\theta_k \in \left\{0,1\right\}$ is the choice between a limit or a deductible being implemented. Here, we set $\theta_k = 0$ if an incident-specific limit is placed, and $\theta_k = 1$ if a deductible is implemented instead;  
    \item $\bm{d} = (d_1,d_2, \dots, d_K)$ are the amounts of incident-specific limits or deductibles, depending on which of the two is imposed on the coverage for each incident type;
    \item $\alpha$ and $\beta$ are choices of risk tolerance levels of the seller and the buyer, respectively;
    \item \begin{equation}\label{eq:seller_loss}
    L_S(\bm{d}, \bm{\theta}; X, O) = \sum_{k = 1}^K\Big(\theta_k\left(X-d_k\right)_+ + (1-\theta_k)\left(X - \left(X-d_k\right)_+\right)\Big)\mathbb{I}_{\{O=k\}},
    \end{equation}
    % $L_S(\bm{d}, \bm{\theta}; X, O) = \sum_{k = 1}^K\Big(\theta_k\left(X-d_k\right)_+ + (1-\theta_k)\left(X - \left(X-d_k\right)_+\right)\Big)\mathbb{I}_{\{O=k\}}$, 
    and 
    \begin{equation}\label{eq:buyer_loss}
      {L_B(\bm{d}, \bm{\theta}; X, O) = \sum_{k=1}^K\Big((1-\theta_k)\left(X-d_k\right)_+ + \theta_k\left(X - \left(X-d_k\right)_+\right)\Big)\mathbb{I}_{\{O=k\}}},
\end{equation}
    % $  {L_B(\bm{d}, \bm{\theta}; X, O) = \sum_{k=1}^K\Big((1-\theta_k)\left(X-d_k\right)_+ + \theta_k\left(X - \left(X-d_k\right)_+\right)\Big)\mathbb{I}_{\{O=k\}}}$ 
    are the seller's and the buyer's loss random variables, respectively, after agreeing with an insurance contract.
\end{itemize}

To solve Problem \eqref{eq:varobj} for Pareto optimal $\bm{\theta}$ and $\bm{d}$, the probability distributions of $L_S$ and $L_B$ in \eqref{eq:seller_loss} and \eqref{eq:buyer_loss} are necessary model inputs to calculate the VaRs in the objective function; these are discussed in Section \ref{sec:model_inputs}. Though Problem \eqref{eq:varobj} is of finite-dimensional in $\bm{d}$, it is also of combinatorial type such that the number of combinations in $\bm{\theta}$ grows exponentially in $K$, making the minimization problem not mathematically tractable; see the number of cases to consider in Proposition 3.1 of \citet{Asimit2021} which seeks for explicit optimal indemnities even if $K=2$. Section \ref{sec:cme} discusses a numerical method and its algorithm to solve the combinatorial optimization problem.

\section{Model Inputs} \label{sec:model_inputs}
The distributions of the seller's and the buyer's loss random variables in \eqref{eq:seller_loss} and \eqref{eq:buyer_loss} belong to the mixture class. Therefore, model inputs to solve Problem \eqref{eq:varobj} are (i) the index set $\mathcal{K}$ of insurable cyber incident types, (ii) the probabilities, $\bm{p} = (p_1, p_2, \dots, p_K)$, of an incident being in an individual category, where $p_k = {\mathbb{P}(O=k)}$, for $k\in\mathcal{K}$ (note that, due to the mutual exclusivity of incident types, $\sum_{k=1}^K p_k = 1$), and (iii) the incident-specific severity of the ground-up loss $(X)\mathbb{I}_{\{O=k\}}$, for $k\in\mathcal{K}$.

% In order to solve Problem \eqref{eq:varobj} for $\bm{d}$ and $\bm{\theta}$, several model inputs are required, including the number of incident types $K$, the probabilities of an incident being in individual categories $\bm{p} = (p_1, p_2, \dots, p_K)$, where $p_k = {\mathbb{P}(O=k)}$, for $k = 1,\dots, K$, and the incident-specific loss severity without the insurance indemnity $(X)\mathbb{I}_{\{O=k\}}$, for $k = 1,2,\dots, K$. Note that due to the mutual exclusivity of incident types, $\sum_{k=1}^K p_k = 1$.

A dataset from Advisen consisting of historical cyber incidents is used to obtain the inputs to determine the loss random variables. 
% To obtain those inputs, a dataset of historical cyber incidents from Advisen is used.
In the dataset, there are 103,061 historical cyber incidents. Its earliest observation dates back to 1987, and the latest observation was recorded in September 2018. This dataset has 125 explanatory variables coded for each incident, which can generally be categorized into four groups as follows: (i) nature of the incident, (ii) victim company information, (iii) consequences of the incident, and (iv) information from any associated lawsuits. With the information on incident types and loss amounts, we can build statistical models to predict the probability vector $\bm{p}$ and the conditional severity distribution of $(X)\mathbb{I}_{\{O=k\}}$, for $k\in\mathcal{K}$. The data and models shall be further elaborated on in the following subsections. 

\subsection{Types of cyber incident}
This paper focuses on the cyber incidents that took place in the US, which represent close to 80\% of all observations. Based on the country information provided for each incident in the dataset, we extract a sample of all US-based incidents, and the sample size is 84,938. 

The response variable is the cyber incident type. The different incident types are the following: (1) cyber extortion, (2) data - malicious breach, (3) data - physically lost or stolen, (4) data - unintentional disclosure, (5) denial of service (DDOS)/system disruption, (6) digital breach/identity thief, (7) identity - fraudulent use/account access, (8) industry controls \& operations, (9) IT - configuration/implementation errors, (10) IT - processing errors, (11) network/website disruption, (12) phishing, spoofing, social engineering, (13) privacy - unauthorized contact or disclosure, (14) privacy - unauthorized data collection, and (15) skimming, physical tampering. Similar to \citet{Kesan2020}, in this paper, these 15 categories are grouped into four types, which are (1) data breach, (2) fraud and extortion, (3) IT error, and (4) privacy violation. The following provides a brief description of each type of cyber incident classified and, together with the ``other'' level, Table \ref{tab:category_size} shows the number of observations for each type of cyber incident classified. Abbreviations in parentheses will occasionally be adopted throughout this paper for clear exposition.

\begin{itemize}
\item \textbf{Privacy violation (PV)} incidents occur when companies collect or disclose individuals' sensitive information, such as personally identifiable information and financial information, without receiving consent from those individuals.
\item \textbf{Data breaches (DB)} are incidents in which data storage devices are breached, causing a possible leakage of confidential information. These incidents can be caused by either hacking activities or the loss of physical devices.
\item Cyber incidents in the \textbf{fraud and extortion (FE)} category are similar to traditional frauds and extortion events, but take place in cyberspace. Cyber frauds typically have forged digital identity involved, such as phishing attacks, and commonly in cyber extortion events, information and information systems are held hostage by intruders for financial gain.
\item In \textbf{IT error (ITE)} events, there is no malicious intent involved. They are caused by incorrectly configured or operating IT systems. 
\end{itemize}

\begin{table}[htbp]
\centering
\begin{tabular}{@{}lr@{}}
\toprule
\textbf{Incident Categories}  & \textbf{Number of Observations} \\ \midrule
(PV) Privacy Violation             & 51315            \\
(DB) Data Breach                   & 26492            \\
(FE) Fraud and Extortion               & 4464             \\
(ITE) IT Error                      & 2102             \\
Other                         & 565             \\ \bottomrule
\end{tabular}
\caption{Number of historical cyber incidents in each category.}
\label{tab:category_size}
\end{table}

\subsection{Incident type occurrence probabilities}

Company-specific information is useful when estimating the occurrence probabilities of different incident types. For example, it is reasonable to assume that technology companies with lots of customer data are more likely to experience data breaches than manufacturing companies, whereas manufacturers with complex industrial control systems are more prone to IT errors that cause disruptions than technology companies. Therefore, we use company characteristics such as industry as predictors for the occurrence probabilities. In this subsection, we present several predictive models that we experimented with for this purpose and make comparisons among them. This approach is similar to insurance rate-making which determines insurance premiums based on the risk characteristics of the insurance buyers.

% In rate-making, insurance buyers are priced based on their own risk characteristics. For example, it is a reasonable assumption that technology companies and manufacturing companies would mostly experience different types of cyber incidents. Therefore, company information is useful when estimating the occurrence probability of each of the four incident types. Various statistical learning approaches are used to establish the relationship between company-and-incident characteristics and those occurrence probabilities. Details of the chosen explanatory variables and the selected predictive models will be given below.

\subsubsection*{Selected explanatory variables}
Among all explanatory variables available in this dataset, we desire manageable and meaningful ones, and thus the variable selection process incorporates the following considerations. 
\begin{itemize}
\item The values of some variables can only be observed after the occurrence of an incident, at which point the type of incident is already known. For example, the settlement cost of a lawsuit associated with an incident is not observable when the incident occurs, and thus this cost has no impact on the type of the incident. Such variables are ruled out from the models. 

% Since the relationship between cyber incident type and variables that are outcomes of a cyber incident is certainly not causal, these variables, such as the various types of losses caused by the incident and the lawsuit information, are discarded.
\item Categorical explanatory variables which have a large number of levels and are downstream in the hierarchy, if any, are removed. For example, among the explanatory variables that represent the geographical information of a company, the one that contains city information is removed, but the one with state information is kept. Likewise, explanatory variables that are more granular industry classification codes, such as Standard Industrial Classification (SIC) codes with more than 2 digits, are removed. This helps reduce the dimensionality of the explanatory variable space when categorical explanatory variables need to be dummified in the modeling process. 
\item For each of the remaining categorical explanatory variables, categories that have too few observations are combined. For example, for the explanatory variable STATE, which has the state information of the victim company, all states that are associated with fewer than 1,000 observations are combined into one category, named ``Other''. This procedure helps mitigate the possible problem that those small categories become absent in training set after the train-test split. Moreover, with fewer categories, the dimensionality of the sample is more manageable after the categorical explanatory variables are dummified. 
\item Auxiliary information, such as the IDs of companies and incidents, provides no predictive power and thus is removed. 
\end{itemize}
As a result, there are 8 explanatory variables remaining. Table \ref{tab:summary_stat} shows the summary statistics of all these explanatory variables; their descriptions are provided in Appendix \ref{append_explan}.

To provide an overview, the processed and cleaned sample contains 84,938 observations of cyber incidents, with the cyber incident type as the multi-class categorical response variable, and 8 explanatory variables, of which 3 are numerical, and 5 are categorical. For categorical explanatory variables, since many of them, such as state and industry, have more than two levels and are non-ordinal, they are numerically coded by dummification. In addition, the three numerical explanatory variables are highly right-skewed, as suggested by their summary statistics, which are expected to have a negative impact on the performance of some linear models that will be tested. Therefore, each of the numerical explanatory variables is log-transformed. For testing and comparing the performance of different models, 30\% of all observations are selected at random and used as the holdout sample, while the remaining 70\% are for training classifiers. 
\afterpage{ % prevent page break and white space caused by landscape table
\begin{landscape}
\begin{table}[t]
\centering
\begin{tabular}{@{}lrrrrrr@{}}
\toprule
\textbf{Explanatory Variable}         & \multicolumn{6}{c}{\textbf{Summary Statistics}}                                                       \\ \midrule
\textit{Numerical}        & \textit{Min.} & \textit{1stQu.} & \textit{Median} & \textit{Mean} & \textit{3rdQu.}   & \textit{Max.} \\ \cmidrule(r){1-1}
EMP                       & 0.0           & 18.0            & 130.0           & 14098.0       & 2120.0            & 2768886.0     \\
log(EMP+1)                & 0.0           & 2.9             & 4.9             & 5.3           & 7.7               & 14.8          \\
NCASE                     & 0.0           & 0.0             & 47.0            & 5338.0        & 1024.0            & 242599.0      \\
log(NACASE+1)             & 0.0           & 0.0             & 3.9             & 4.0           & 6.9               & 12.4          \\
REV                       & 0.0           & 3.2             & 31.2            & 6629.2        & 617.5             & 496785.0      \\
log(REV+1)                & 0.0           & 1.4             & 3.5             & 4.2           & 6.4               & 13.1          \\ \midrule
\textit{Categorical}      & \multicolumn{6}{c}{\textit{Levels}}                                                                   \\ \cmidrule(r){1-1}
MON                       & JAN           & MAR             & APR             & FEB           & (8 other levels)  & Missing       \\
\multicolumn{1}{r}{count} & 23101         & 6182            & 5856            & 5788          & 42592             & 1419          \\
STATE                     & CA            & Other           & NY              & MA            & (20 other levels) & Missing       \\
\multicolumn{1}{r}{count} & 13915         & 11519           & 6835            & 5934          & 46361             & 374           \\
CTYPE                     & PRV           & PUB             & OTHER           & -             & -                 & Missing       \\
\multicolumn{1}{r}{count} & 54584         & 25967           & 4387            & -             & -                 & 0             \\
IND                       & I             & H               & G               & E             & (4 other levels)  & Missing       \\
\multicolumn{1}{r}{count} & 43932         & 22465           & 5897            & 4058          & 8586              & 0             \\
YEAR                      & BEFORE2012    & AFTER2012       & -               & -             & -                 & Missing       \\
\multicolumn{1}{r}{count} & 49282         & 34237           & -               & -             & -                 & 1419          \\ \bottomrule
\end{tabular}%
\caption{Summary statistics of explanatory variables. For each categorical explanatory variable, information is displayed for at most four categories with the most number of observations. Other levels are used for modeling, but are truncated in this table for ease of reading.}
\label{tab:summary_stat}
\end{table}
\end{landscape}}

\subsubsection*{Multi-class classification} \label{sec:classification}
Predicting which type an incident falls into once it occurs is essentially a multi-class classification problem. In this study, we explored several commonly used classification models, including decision trees, random forests, gradient-boosted trees, linear discriminant analysis, multinomial logistic regression, and multi-layer perceptron. In the end, we build a stacking classifier on top of these models for any possible performance gains. Because all these methods are well established and documented in a broad volume of literature, we shall only provide brief and mostly qualitative overviews of them in Appendix \ref{append:classifiers} and focus instead on their performance and prediction results. % to add citations for each method

\subsubsection*{Model comparisons}
Table \ref{tab:category_size} suggests that the four incident categories are unbalanced, and a large part of them are within the types of data breach and privacy violation types. In this circumstance, several metrics, as summarized in \citet{Grandini2020}, could be used to evaluate model performance. In this paper, we shall report the balanced accuracy of each model as a proxy of its performance. The balanced accuracy score is defined as follows:
\begin{equation}
\begin{aligned}
\text{Balanced Accuracy} =  \frac{1}{2K}\Bigg(\sum_{k=1}^K \Bigg(&\;  \frac{\sum_{i=1}^N \mathbb{I}_{\{O_i = \hat{O}_i = k\}}}{\sum_{i=1}^N \mathbb{I}_{\{O_i = \hat{O}_i = k\}} + \sum_{i=1}^N \mathbb{I}_{\{O_i = k \ne \hat{O}_i\}}} \\ & + \frac{\sum_{i=1}^N \mathbb{I}_{\{O_i = \hat{O}_i \ne k\}}}{\sum_{i=1}^N \mathbb{I}_{\{O_i = \hat{O}_i \ne k\}} + \sum_{i=1}^N \mathbb{I}_{\{\hat{O}_i = k \ne O_i\}}}\Bigg) \Bigg),
\end{aligned}
\label{eq:bal_acc}
\end{equation}
where
\begin{itemize}
    \item $N$ is the size of the test set;
    \item $O_i$ and $\hat{O}_i$ are the observed and the predicted occurring incident types of record $i$, respectively, for $i = 1,2,\dots, N$.
\end{itemize}

The one-versus-all metric evaluates the model performance by comparing the predictions in one class against those in the combination of all other classes. Specifically, for each $k\in\mathcal{K}$, $\sum_{i=1}^N \mathbb{I}_{\{O_i = \hat{O}_i = k\}}$ is the number of true positives with respect to incident type $k$, $\sum_{i=1}^N \mathbb{I}_{\{O_i = k \ne \hat{O}_i\}}$ is the false negative count, and $\sum_{i=1}^N \mathbb{I}_{\{O_i = \hat{O}_i \ne k\}}$ and $\sum_{i=1}^N \mathbb{I}_{\{\hat{O}_i = k \ne O_i\}}$ correspond to true negatives and false positives, respectively. Therefore,  $\frac{1}{2}\Bigg(  \frac{\sum_{i=1}^N \mathbb{I}_{\{O_i = \hat{O}_i = k\}}}{\sum_{i=1}^N \mathbb{I}_{\{O_i = \hat{O}_i = k\}} + \sum_{i=1}^N \mathbb{I}_{\{O_i = k \ne \hat{O}_i\}}}  + \frac{\sum_{i=1}^N \mathbb{I}_{\{O_i = \hat{O}_i \ne k\}}}{\sum_{i=1}^N \mathbb{I}_{\{O_i = \hat{O}_i \ne k\}} + \sum_{i=1}^N \mathbb{I}_{\{\hat{O}_i = k \ne O_i\}}}\Bigg)$ represents the $k$-th class-specific balanced accuracy, and Equation \eqref{eq:bal_acc} represents the average of these balanced accuracy values across all classes.

After all the classifiers are tuned and trained, their performance in terms of the balanced accuracy of the predictions in the test set (holdout sample) is presented in Table \ref{tab:model_comparison}. The table shows the by-class balanced accuracy of each classifier as well as the average across all classes, and the best score is highlighted in each class or among all averages.

\begin{table}[htbp]
\centering
\begin{tabular}{@{}lrrrrr@{}}
\toprule
\multirow{2}{*}{\textbf{Classifier}} & \multicolumn{4}{r}{\textbf{By-Class Balanced   Accuracy}}             & \multirow{2}{*}{\textbf{Average}} \\
                                     & \textit{PV}     & \textit{DB}     & \textit{FE}     & \textit{ITE}    &                                   \\ \midrule
Decision tree                        & 0.8373          & 0.7827          & 0.6369          & 0.6357          & 0.7231                            \\
Random forest                        & 0.8506          & 0.7976          & 0.7151          & 0.7193          & 0.7707                            \\
Gradient boosted trees      & 0.8534          & 0.7993          & 0.7270          & 0.6991          & 0.7697                            \\
Linear discriminant analysis         & 0.7005          & 0.6777          & 0.5244          & 0.5746          & 0.6193                            \\
Multinomial logistic regression      & 0.7054          & 0.6817          & 0.4744          & 0.4864          & 0.5870                            \\
Multi-layer perceptron                       & 0.7515          & 0.7138          & 0.8571          & 0.4864          & 0.7022                            \\
Stack (Trees)                        & 0.8592          & 0.8021          & 0.7823          & \textbf{0.7799} & \textbf{0.8059}                   \\
Stack (Linear classifiers excluded)                    & \textbf{0.8622} & 0.8030          & \textbf{0.7865} & 0.7698          & 0.8054                            \\
Stack (All)                          & 0.8616          & \textbf{0.8032} & 0.7760          & 0.7789          & 0.8049                            \\ \bottomrule
\end{tabular}
\caption{Comparison of classification models.}
\label{tab:model_comparison}
\end{table}

To summarize the performance comparison among different models, it is easy to observe that stacking models overall outperform their individual base classifiers. The best by-class accuracy scores and the best average accuracy score are all achieved by stacking models. Second, the distinction between the three stacking models is small. This suggests that the inclusion of classifiers other than tree-based models does not substantially improve the model performance. In addition, the by-class accuracy scores and the average balanced accuracy score attained by stacking models are generally close to $80\%$, which is a reasonably large proportion. Based on these three observations, the stacking classifier built on top of the three tree-based classifiers is adopted to predict incident-type occurrence probabilities.

The numerical results, in terms of the predicted probabilities of the four incident types given different sets of company and incident characteristics, are presented in Table \ref{tab:fa_num_results} of Section \ref{sec:fun_approx}.

\subsection{Incident-specific loss severity} \label{sec:severity_modeling}
After predicting the occurrence probability of individual incident type, the next step is to model the loss severity conditioning on a given incident type. Therefore, another useful attribute of the dataset is the loss information on each incident. Excluding observations that belong to the ``other'' category, among all 84,373 incidents that belong to one of the 4 cyber incident categories, there are 3,978 observations with known losses. The summary statistics of those losses are shown in Table \ref{tab:loss_summary}.

\begin{table}[htbp]
\centering
\begin{tabular}{@{}lrrrrrrr@{}}
\toprule
\textbf{Type} & \textbf{Count} & \textbf{Min.} & \textbf{1st Qu.} & \textbf{Median} & \textbf{Mean} & \textbf{3rd Qu.} & \textbf{Max.}    \\ \midrule
PV  & 2,016 & 30   & 4,838   & 28,572  & 4,606,605  & 501,556   & 1,000,000,000 \\
DB  & 610   & 1    & 18,195  & 202,500 & 11,835,168 & 1,677,500 & 4,000,000,000 \\
FE  & 1,137 & 180  & 10,025  & 137,817 & 11,884,945 & 1,700,000 & 1,750,000,000 \\
ITE & 215   & 200  & 20,000  & 194,850 & 4,994,540  & 1,050,000 & 200,000,000   \\
All & 3,978 & 1    & 6,511   & 60,000  & 7,112,494  & 945,287   & 4,000,000,000 \\ \bottomrule
\end{tabular}
\caption{Summary statistics of losses by incident type.}
\label{tab:loss_summary}
\end{table}

Prior to fitting any reasonable distributions to each conditional loss, it is essential to realize that the conditional loss distributions could be significantly different among cyber incident types. Table \ref{tab:ks_table} shows the pairwise two-sample Kolmogorov-Smirnov (KS) tests on empirical loss distributions among various incident types. The numbers below the diagonal are the $p$-values of these KS tests. Based on the significance level $0.05$, a $p$-value lower than this level suggests that the null hypothesis, that the two empirical samples being compared are from the same underlying distribution, should be rejected. Due to symmetry, the statistical decisions are noted above the diagonal. The result shows that the difference among the conditional losses of the four incident types is statistically significant, and hence there is the necessity of individually modeling each incident type's loss severity, justifying the incident-specific insurance coverage.

\begin{table}[h]
\centering
\begin{tabularx}{0.7\textwidth}{lRRRR}
\toprule
    & PV              & DB        & FE                  & ITE             \\ \midrule
PV  & -        &\textbf{reject}  &\textbf{reject}             &\textbf{reject}       \\
DB  &  0.00E+00 & -  & \textbf{reject}     & \textbf{reject} \\
FE  & 0.00E+00 & 0.00E+00  & -            & \textbf{reject} \\
ITE &  2.83E-05  & 1.70E-06  & 2.84E-05            & -        \\
\bottomrule
\end{tabularx}
\caption{Pairwise comparison between empirical distributions of different incident types using pairwise two-sample Kolmogorov-Smirnov test.}
\label{tab:ks_table}
\end{table}

In addition, to show that loss severity distributions are mainly conditional on incident types but not on additional explanatory variables, we performed an ANOVA test with two linear regression models. One is a full model with total loss as the response variable, and with incident type, as well as all explanatory variables summarized in Table \ref{tab:summary_stat}. The other model is a simple linear regression, and it has the total loss as the response variable, but has incident type as its only explanatory variable. The ANOVA test result, presented in Table \ref{tab:anova}, shows that the $p$-value of the test, $0.1447$, is higher than any common choices of significance level, \textit{e.g.}, $0.05$, $0.01$, etc. Therefore, we can conclude that the additional explanatory variables in Table \ref{tab:summary_stat} for the full model contribute little to the explanation of the variance in losses. This lack of explanatory power could be an artifact of the limited data size, which makes it infeasible to build company-specific severity models, but this situation can potentially be improved by a growing number of cyber incident records.

\begin{table}[h]
\centering
\begin{tabular}{@{}lrr@{}}
\toprule
                           & \multicolumn{1}{r}{Full} & \multicolumn{1}{r}{Simple} \\ \midrule
Residual   Degrees of Freedom    & $2783$                            & $2829$                                          \\
Residual Sum of Squares         & $2.33\times 10^{19}$              & $2.38 \times 10^{19}$                           \\
Difference in Degrees of Freedom &                                   & $-46$                                           \\
Difference in Sum of Squares     &                                   & $-4.71\times 10^{17}$                             \\
F Statistic                     &                                   & $1.2238$                                        \\
$p$-value                         &                                   & $0.1447$                                        \\ \bottomrule
\end{tabular}
\caption{ANOVA test between the full model and the model with only incident type as its explanatory variable.}
\label{tab:anova}
\end{table}

\begin{figure}[ht]
     \centering
     \begin{subfigure}[b]{0.75\textwidth}
    \centering
    \includegraphics[width=\textwidth]{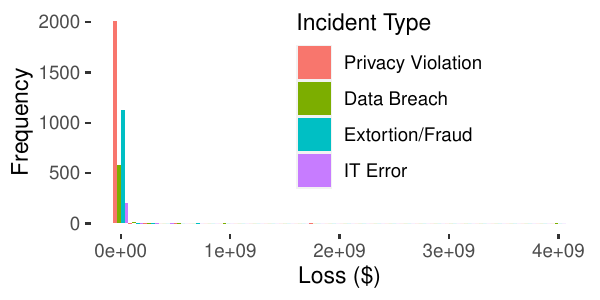}
    \caption{All losses.}
    \label{fig:loss_hist_all}
     \end{subfigure}
     \hfill
     \begin{subfigure}[b]{0.75\textwidth}
    \centering
    \includegraphics[width=\textwidth]{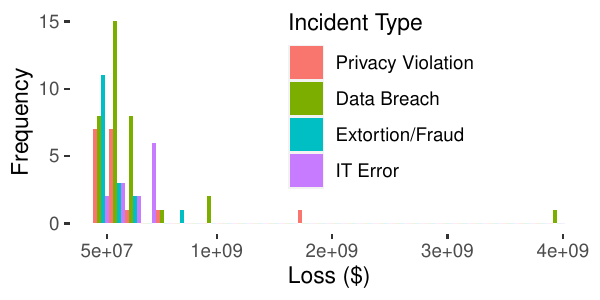}
    \caption{Losses greater than or equal to \$10 million.}
    \label{fig:loss_hist_zoom}
     \end{subfigure}
        \caption{Histogram of losses of different incident types.}
        \label{fig:loss_hist}
\end{figure}

Figure \ref{fig:loss_hist} shows that the majority of the cases realize moderately small losses, and a small number of cases realize substantial losses, indicating that the conditional loss distributions of all incident types are likely to be right-skewed. Therefore, we explore well-known distributions that could capture such a skewness, namely log-normal, exponential, gamma, and Weibull, for each cyber incident type.

The parameters of each distribution are fitted using the maximum likelihood estimation method. All fitted distributions for the same cyber incident type are then compared based on their Akaike Information Criterion (AIC), which is defined as
\begin{equation*}
\text{AIC} = 2\hat{N} - 2\ln(\hat{L}),
\end{equation*}
where $\hat{N}$ is the number of parameters to be estimated, and $\hat{L}$ is the maximum likelihood of the fitted model; a model with a smaller AIC value is better. Table \ref{tab:fit_summary} summarizes the best-fitted loss distributions for all cyber incident types. A comparison of the AICs of different fitted distributions is presented in Table \ref{tab:fitted_aics} of Appendix \ref{append_aics}. The distribution fitting results suggest that it is most appropriate to treat losses of different incident types as log-normal random variables with different log-mean and log-standard-deviation parameters, as given in Table \ref{tab:fit_summary}. 
% Additional plots for evaluating the goodness-of-fit of the fitted log-normal distributions are relegated to Appendix \ref{append_gof_plots}.

\begin{table}[h]
\centering
\begin{tabular}{@{}llrr@{}}
\toprule
Type & Distribution & $\mu$   & $\sigma$ \\ \midrule
PV   & log-normal   & -2.5996 & 3.2798   \\ 
DB   & log-normal   & -0.7916 & 3.1122   \\
FE   & log-normal   & -3.4100 & 2.8577   \\
ITE  & log-normal   & -1.9557 & 3.3629   \\
\bottomrule
\end{tabular}
\caption{Best fitted loss distribution of each incident type.}
\label{tab:fit_summary}
\end{table}

\subsection{Section summary}
Thus far, all the necessary model inputs for solving Problem \eqref{eq:varobj} are obtained based on a dataset of historical cyber incidents. A brief recapitulation of the key findings in this section is as follows:
\begin{itemize}
    \item In this study, the incident-specific cyber insurance policy would provide coverage for four types of incidents, including privacy violation, data breach, cyber fraud and extortion, and IT errors, \textit{i.e.}, $K=4$.
    \item The occurrence probabilities of individual incident types are specific to each company and incident, and the probability vector $\bm{p} = (p_1, p_2, \dots, p_K)$ is obtained by feeding the selected company and incident features into a trained stacking classifier. This classifier achieves a balanced accuracy score of around 80\%, as defined in Equation \eqref{eq:bal_acc}, for each incident type, which is superior to all other predictive models surveyed (see Table \ref{tab:model_comparison}).
    \item Incident-specific losses are statistically significant to be different in terms of severity. Additional explanatory variables, apart from the incident type, do not seem crucial for modeling the loss severity. Log-normal distributions provide the best fit for all conditional loss severity with estimated parameters presented in Table \ref{tab:fit_summary}.
    %modeled by log-normal random variables with distribution parameters estimated based on historical loss data, and the values of the estimated parameters are 
\end{itemize}
This section demonstrates the feasibility of obtaining the model inputs necessary to solve Problem \eqref{eq:varobj} from real-world data. With approximately 80\% balanced accuracy scores associated with predictions of incident types and well-fitted condition loss severity distributions, they are expected to be practical and reliable to proceed to the next step.

\section{Solution by Cross-Entropy Method} \label{sec:cme}
As discussed before, even with the necessary model inputs to solve Problem \eqref{eq:varobj}, despite the minimization problem being finite-dimensional, it is also combinatorial and, therefore, cannot be solved analytically to derive the Pareto optimal $\bm{\theta}$ and $\bm{d}$, which are denoted as $\bm{\theta}^*$ and $\bm{d}^*$. 
%Given model inputs, Problem \eqref{eq:varobj} can be solved. It is combinatorial with discrete choices of indemnity types, and the number of cases that should be considered grows exponentially with the number of covered incident types, which makes the analytical solution impractical to derive.
In this paper, we resort to the cross-entropy method (CEM) proposed in \citet{Rubinstein1999, Rubinstein2001}. The following subsection provides a brief review of the CEM adopting our own minimization problem in \eqref{eq:varobj}.

\subsection{Brief review of cross-entropy method}
To simplify the notations, let $\mathbf{z}=\left(\bm{d}, \bm{\theta}\right)\in\mathcal{Z}=\mathbb{R}_+^4\times\left\{0,1\right\}^4$, and simply write the objective function in Problem \eqref{eq:varobj} as $F\left(\mathbf{z}\right)$ (keeping in mind that it depends on the (conditional) distributions of $X$ and $O$ which are model inputs from Section \ref{sec:model_inputs}). Thus, Problem \eqref{eq:varobj} is $\min_{\mathbf{z}\in\mathcal{Z}}F\left(\mathbf{z}\right)$. It is well-known that an $\mathbf{z}^*\in\mathcal{Z}$ is optimal, if and only if, for any $\varepsilon>0$, there exists an $\mathbf{z}\in\mathcal{Z}$ such that $F\left(\mathbf{z}\right)<F\left(\mathbf{z}^*\right)+\varepsilon$; this is also equivalent with requiring that for any $\varepsilon>0$, $\mathbf{z}^*\in\mathcal{Z}_{\varepsilon}\left(\mathbf{z}^*\right)$, where $\mathcal{Z}_{\varepsilon}\left(\mathbf{z}^*\right)=\left\{\mathbf{z}\in\mathcal{Z}:F\left(\mathbf{z}\right)<F\left(\mathbf{z}^*\right)+\varepsilon\right\}$.

The CEM is a numerical method to solve Problem \eqref{eq:varobj} by adopting a probabilistic approach to the characterization for optimality above. To this end, let $\mathbf{Z}$ be a random vector with range $\mathcal{Z}$. The optimality can then be further characterized by the fact that considering all possible distributions of $\mathbf{Z}$, for any $\varepsilon>0$, $\mathbb{P}\left(\mathbf{Z}\in\mathcal{Z}_{\varepsilon}\left(\mathbf{z}^*\right)\right)=\mathbb{P}\left(F\left(\mathbf{Z}\right)<F\left(\mathbf{z}^*\right)+\varepsilon\right)$ is maximized by the Dirac delta distribution $\delta_{\mathbf{z}^*}$ with the optimal objective value being $1$ uniformly for all $\varepsilon>0$. However, to maximize this associated probabilistic objective, $\mathbf{z}^*$, the unknown to be solved for, has to be known ex-ante. Hence, together with considering only parametric distributions of $\mathbf{Z}$, the CEM solves the approximation counterpart of $\mathbf{z}^*$ as follows: for any $\tau$ being larger but close enough to $F\left(\mathbf{z}^*\right)$, solve an $\mathbf{u}^*\left(\tau\right)\in\mathcal{U}$ which maximizes $\mathbb{P}\left(F\left(\mathbf{Z}\right)\leq\tau;\mathbf{u}\right)$, where $\mathbf{Z}$ follows a distribution with a parametric density function $f\left(\mathbf{z};\mathbf{u}\right)$, for $\mathbf{z}\in\mathcal{Z}$ and $\mathbf{u}\in\mathcal{U}$, and $\mathcal{U}$ is the parameter set for the distributions of $\mathbf{Z}$ under consideration. But still, without knowing $\mathbf{z}^*$, the level $\tau$ in this approximation counterpart could not be chosen appropriately.

The CEM proposes a two-phase approach to construct a sequence $\left\{\left(\hat{\tau}_t,\hat{\mathbf{u}}_t\right)\right\}_{t=1}^{\infty}$ such that it will converge to an $\left(\hat{\tau},\hat{\mathbf{u}}\right)$, where $\left(\hat{\tau},\hat{\mathbf{u}}\right)$ is close to $\left(F\left(\mathbf{z}^*\right),\mathbf{u}^*\right)$, and thus serves as  for the unknown $\left(F\left(\mathbf{z}^*\right),\mathbf{u}^*\right)$, and where $\mathbf{u}^*\in\mathcal{U}$ such that $f\left(\cdot;\mathbf{u}^*\right)$ is the corresponding Dirac density of the Dirac delta distribution $\delta_{\mathbf{z}^*}$. The main steps of the two-phase construction are as follows. Initialize an $\hat{\mathbf{u}}_0\in\mathcal{U}$, and let $t$ iterate through $1,2,\dots$ with $\left(\hat{\tau}_t,\hat{\mathbf{u}}_t\right)$ being updated in each iteration. The $\hat{\tau}_t$, for $t=1,2,\dots$, is defined as the $\varrho$-th quantile of the sample drawn from the distribution $f\left(\cdot;\hat{\mathbf{u}}_{t-1}\right)$, where $\varrho\in\left(0,1\right)$ is close to $0$ but is moderately small and uniform for all $t=1,2,\dots$; in turn, the $\hat{\mathbf{u}}_t$, for $t=1,2,\dots$, is defined as the maximizer of the following problem:
\begin{equation}
\max_{\mathbf{u}_t\in\mathcal{U}}\hat{\mathbb{E}}\left[\mathbb{I}_{\left\{F\left(\mathbf{Z}\right)\leq\hat{\tau}_t\right\}}\ln f\left(\mathbf{Z};\mathbf{u}_t\right);\hat{\mathbf{u}}_{t-1}\right],
\label{eq:cross-entropy}
\end{equation}
where $\hat{\mathbb{E}}\left[\cdot;\hat{\mathbf{u}}_{t-1}\right]$ represents the estimated mean of the same sample drawn from the distribution $f\left(\cdot;\hat{\mathbf{u}}_{t-1}\right)$. This two-phase update repeats itself until the estimated variance of the next sample drawn from the updated distribution $f\left(\cdot;\hat{\mathbf{u}}_{t}\right)$ is small enough and the $\hat{\tau}_t$ does not change from those in the last few update steps (see \cite{Benham2017}). Therefore, the main idea of the construction is, first estimating an unknown level $\hat{\tau}_t$ being larger but close enough to $F\left(\mathbf{z}^*\right)$ as it is the $\varrho$-th sample quantile with $\varrho$ being close to $0$ using the parameter $\hat{\mathbf{u}}_{t-1}$, and second, estimating a parameter $\hat{\mathbf{u}}_t$ such that the event $\left\{F\left(\mathbf{Z}\right)\leq\hat{\tau}_t\right\}$ with $\mathbf{Z}\sim f\left(\cdot;\hat{\mathbf{u}}_{t}\right)$ is much more likely to happen, compared to the case that $\mathbf{Z}\sim f\left(\cdot;\hat{\mathbf{u}}_{t-1}\right)$ when its likelihood is given by a moderately small $\varrho$, so to sequentially maximize $\mathbb{P}\left(F\left(\mathbf{Z}\right)\leq\hat{\tau}_t;\mathbf{u}_{t}\right)$. In the second phase, the estimated parameter $\hat{\mathbf{u}}_t$ is to minimize the {\it cross-entropy} between the unknown optimal density, for importance sampling, and a reference density in the same parametric family; this is equivalent to maximizing \eqref{eq:cross-entropy}.

Note that in solving \eqref{eq:cross-entropy}, only the observations in the sample drawn from the distribution $f\left(\cdot;\hat{\mathbf{u}}_{t-1}\right)$ satisfying the event condition, $F\left(\mathbf{Z}\right)\leq\hat{\tau}_t$, are useful to calculate its objective; this motivates the introduction of an elite sample consisting of observations in the original sample satisfying the condition. Algorithm \ref{alg:cme} provides the pseudo-code of the CEM.

\SetKwComment{Comment}{//}{}
\SetKw{Or}{or}
\SetKw{Return}{return}
\begin{algorithm}
\SetNoFillComment
\caption{Cross-Entropy Method}\label{alg:cme}
\KwIn{
$\varrho \in (0, 1)$ \Comment{Elite sample proportion}\\
$f\left(\cdot;\cdot\right)$ \Comment{Parametric distribution of $\mathbf{Z}$}\\
$\hat{\mathbf{u}}_0$ \Comment{Initial parameter for $f\left(\cdot;\cdot\right)$} \\
$\nu$ \Comment{Variance stopping criterion} \\
$l$ \Comment{Lag between iterations in which approximated optimums are compared}
$N$ \Comment{Sample size}
$F$ \Comment{Objective function}
}
\KwOut{Sample from $f\left(\cdot;\hat{\mathbf{u}}\right)$ \Comment{Approximation of $\mathbf{z}^*$}}
$t \gets 0$

$N_E \gets \lceil \varrho N \rceil$ \Comment{Elite sample size}

Draw random sample $\left\{\mathbf{z}_1^{\left(t\right)}, \mathbf{z}_2^{\left(t\right)}, \dots, \mathbf{z}_N^{\left(t\right)}\right\}$ from distribution with density $f(\cdot; \hat{\mathbf{u}}_t)$

Calculate element-wise sample variances $V^{\left(t\right)}_k$, for $k = 1,2,\dots,2K$

$\Delta \gets \infty$

\While{$\max\left(V^{\left(t\right)}_1, V^{\left(t\right)}_2, \dots, V^{\left(t\right)}_{2K}\right) > \nu$ \Or $\Delta \ne 0$ }{
$y^{\left(t\right)}_i \gets F(\mathbf{z}_i^{\left(t\right)})$ for $i = 1, 2, \dots, N$

Select elite sample $\mathcal{E}^{\left(t\right)}=\left\{\mathbf{z}_i^{\left(t\right)}:F\left(\mathbf{z}_i^{\left(t\right)}\right)=y_{(i)}^{\left(t\right)},\;i=1,2,\dots,N_E\right\}$
%\left(y_{(1)}, y_{(2)}, \dots, y_{(N_E)}\right)$

$\hat{\tau}_t \gets y^{\left(t\right)}_{(N_E)}$

\If{$t \ge l$}{
    $\Delta \gets \max(|\hat{\tau}_t - \hat{\tau}_{t-1}|, |\hat{\tau}_{t} - \hat{\tau}_{t-2}|, \dots, |\hat{\tau}_{t} - \hat{\tau}_{t-l}|)$
}

Obtain an estimate $\hat{\mathbf{u}}_{t+1}$ using the elite sample $\mathcal{E}^{\left(t\right)}$  

$t \gets t+1$

Draw random sample $\left\{\mathbf{z}_1^{\left(t\right)}, \mathbf{z}_2^{\left(t\right)}, \dots, \mathbf{z}_N^{\left(t\right)}\right\}$ from distribution with density $f(\cdot; \hat{\mathbf{u}}_t)$

}

$T \gets t$ 

\Return $\mathbf{z}_1^{\left(T\right)}$ \Comment{Arbitrarily choosing the first observation because observations are similar due to small variance}

\end{algorithm}

\subsection{Implementation considerations and results} \label{sec:implement}
The actual implementation of this optimization problem consists of two parts, including computing the objective function and implementing the CEM.
%They are discussed in the section as follows, accompanied by results in Table \ref{tab:optim_results}.

The objective function is the sum of the VaRs of both parties. As suggested by Equations \eqref{eq:seller_loss} and \eqref{eq:buyer_loss}, the distribution of either the buyer's or the seller's loss random variable is a mixture of distributions of incident-specific losses, and the mixture distribution has no closed-form quantile function that returns the Value-at-Risk of concern easily. Therefore, relying on the mixture distribution function, we adopted a lookup approach to numerically find a point that satisfies Equation \eqref{eq:VaR} on a fine grid. In this study, risk tolerance levels of the seller and the buyer are set at $\alpha=0.95$ and $\beta=0.9$, respectively.

Although the CEM has a guaranteed asymptotic convergence (see, for example,  \citet{Rubinstein2002, Margolin2005}), the number of iterations needed to meet the stopping criteria is uncertain because of the randomness introduced during the sampling processes. Therefore, a common approach to addressing this issue in algorithm implementation is adding additional stopping rules to ensure that the running time of optimization tasks is manageable. Typical rules may include enforcing a maximum number of iterations and assuming convergence if the value of the objective function has not experienced a large enough improvement for a predefined number of iterations. These rules, along with the randomness in sampling, cause the algorithm in our implementation not always converge timely and not always stop at the same value. As suggested by \citet{Benham2017}, the optimization process shall be repeated several times for quality investigation and assurance. Hence, for each set of incident probabilities predicted based on certain company and incident characteristics, we ran 50 trials of the CEM to solve the optimization with different random seeds, while holding other specifications constant (see Appendix \ref{append:params}). Results from all trials are stored for further analysis. 

Table \ref{tab:optim_results} shows the results of five out of those 50 trials. Trial 5 converged to an ``optimum'' larger than that achieved by Trial 1, suggesting that the algorithm could possibly stop at a non-optimal point. This highlights the necessity of multiple trials. Another observation is that, although most of the trials reach a converging state after a few numbers of iterations (see, \textit{e.g.}, Trials 1, 4, and 5), there exist initial random states that cause the program to fail to meet the convergence criteria when the preset maximum number of iterations is reached, \textit{e.g.}, Trials 2 and 3. Unless reducing the maximum number of iterations at the risk of more non-convergent trials, the great gap between numbers of iterations needed by different trials brings the challenge that even in a multiprocessing environment, congestion occurs when computational resources are occupied by those long-running trials, making solving the optimization problem with repeating trials time-consuming. 

Table \ref{tab:time_cem} shows the amount of time spent on finding optimal insurance designs for $5$ individual organizations using the CEM. The computation is done on 4 IBM POWER9 CPU cores. For each company, it takes around 2 minutes to complete each trial. Given that we ran $50$ trials for each company, the running time generally could easily exceed an hour. 

\begin{table}[htbp]
\centering
\begin{tabular}{@{}lrrrrr@{}}
\cmidrule(l){2-6}
 & \multicolumn{5}{c}{Organization} \\ \midrule
 & \multicolumn{1}{c}{1} & \multicolumn{1}{c}{2} & \multicolumn{1}{c}{3} & \multicolumn{1}{c}{4} & \multicolumn{1}{c}{5} \\ \midrule
Count & 50 & 50 & 50 & 50 & 50 \\
Mean (s) & 211.13 & 108.06 & 104.13 & 123.65 & 125.21 \\
Standard Deviation (s) & 317.31 & 160.15 & 162.98 & 181.67 & 185.44 \\
Minimum (s) & 24.97 & 11.68 & 9.62 & 15.41 & 14.84 \\
25\% Quantile (s) & 43.58 & 20.27 & 22.62 & 27.85 & 25.94 \\
50\% Quantile (s) & 68.18 & 34.78 & 36.63 & 45.37 & 40.52 \\
75\% Quantile (s) & 155.92 & 86.31 & 83.39 & 99.20 & 94.85 \\
Maximum (s) & 975.61 & 538.80 & 539.01 & 612.82 & 542.56 \\
Sum (s) & 10556.59 & 5402.97 & 5206.36 & 6182.30 & 6260.34 \\ \bottomrule
\end{tabular}
\caption{Descriptive statistics on the number of seconds (s) spent on finding the optimal insurance design using the CEM}
\label{tab:time_cem}
\end{table}

Other than the long running time problem, the solution to the minimization problem in Equation \eqref{eq:varobj} is not unique, as presented in \citet{Asimit2021} for the case of two incident types. The value of each element of $\bm{d}^*$ could lie in some intervals, composed of points that result in the same minimum. Table \ref{tab:optim_results} shows that although both Trial 1 and 4 reached the converging state and attained the same minimum, their solutions, $\bm{d}^*$ in particular, differ. Nevertheless, the results provide common ground to insurers and policyholders on the design of an incident-specific cyber insurance contract. As shown in Table \ref{tab:optim_results}, based on the company's own characteristics and the risk preferences of the company and the insurer, both parties should agree that in comprehensive cyber insurance covering PV, DB, FE, and ITE, a policy limit should be implemented for PV, whereas deductibles should be applied on other incident-specific coverages, since $\bm{\theta}^*=\left(0,1,1,1\right)$. While holding the risk of both parties unchanged, deductible and limit amounts can be negotiated for other possible considerations, one of which is discussed in the next section.

Despite the computational challenges and non-unique solutions, the benefit of such an incident-specific insurance contract is readily seen from the comparison between the risks taken by both parties with and without insurance. In the folloiwng, refer to Trials $1$ to $4$.  If the company chooses not to buy an incident-specific policy specified in Table \ref{tab:optim_results}, it will take the risk of \$13.6984 million. If it chooses to get covered, the appropriate amount of premium $\pi$ (in millions) paid by the buyer should satisfy $\$2.2977 \le \pi \le \$4.6595$ according to Inequality \eqref{eq:premium_range}, and that will yield a risk reduction of \$4.6595 million, which is always greater than the paid premium. From the insurer's perspective, the insurance contract will bring a risk of \$2.2977 million, but that will be fully compensated by the collected premium. As a result, both parties are mutually benefited on their own objectives; their aggregate risk is reduced by \$2.3618 million, from \$13.6984 million to \$11.3366 million, which could be effective if the premium is agreed to be lower than the reduced aggregate risk in its range. Again, note that the premium discussed in this paper is derived economically which could be actuarially priced in the insurance practice.

% \usepackage{graphicx}
\begin{table}[]
\centering
\resizebox{\textwidth}{!}{%
\begin{tabular}{lcrrrr}
\hline
\textbf{Trial} & \multicolumn{1}{r}{\textbf{1}} & \textbf{2} & \textbf{3} & \textbf{4} & \textbf{5} \\ \hline
$p_1^*$ (PV) & \multicolumn{5}{c}{0.3383} \\
$p_2^*$ (DB) & \multicolumn{5}{c}{0.5717} \\
$p_3^*$ (FE) & \multicolumn{5}{c}{0.0700} \\
$p_4^*$ (ITE) & \multicolumn{5}{c}{0.0200} \\ \hline
Random seed & \multicolumn{1}{r}{0} & 8 & 13 & 6 & 2 \\
No. of iterations & \multicolumn{1}{r}{11} & 401 & 401 & 41 & 54 \\
\begin{tabular}[c]{@{}l@{}}Convergence \\ status\end{tabular} & \multicolumn{1}{r}{\begin{tabular}[c]{@{}r@{}}Variance \\ converged\end{tabular}} & \begin{tabular}[c]{@{}r@{}}Not \\ converge\end{tabular} & \begin{tabular}[c]{@{}r@{}}Not \\ converge\end{tabular} & \begin{tabular}[c]{@{}r@{}}Variance \\ converge\end{tabular} & \begin{tabular}[c]{@{}r@{}}Variance \\ converged\end{tabular} \\ \hline
Optimum (millions) & \multicolumn{4}{c}{11.3366} & 11.3381 \\
$\theta_1^*$ (PV) & \multicolumn{4}{c}{0} & 0 \\
$\theta_2^*$ (DB) & \multicolumn{4}{c}{1} & 1 \\
$\theta_3^*$ (FE) & \multicolumn{4}{c}{1} & 1 \\
$\theta_4^*$ (ITE) & \multicolumn{4}{c}{1} & 1 \\
$d_1^*$ (millions, PV) & \multicolumn{1}{r}{0.0531} & 0.0474 & 0.1074 & 0.0511 & 0.0334 \\
$d_2^*$ (millions, DB) & \multicolumn{1}{r}{0.1011} & 0.1153 & 0.0982 & 0.1276 & 0.0941 \\
$d_3^*$ (millions, FE) & \multicolumn{1}{r}{0.1167} & 0.0210 & 0.0107 & 0.0979 & 0.0520 \\
$d_4^*$ (millions, ITE) & \multicolumn{1}{r}{0.1151} & 0.0195 & 0.0435 & 0.0314 & 0.0585 \\ \hline
\begin{tabular}[c]{@{}l@{}}Buyer's risk without \\ insurance (millions)\end{tabular} & \multicolumn{4}{c}{13.6984} & 13.6984 \\
\begin{tabular}[c]{@{}l@{}}Buyer's risk with \\ insurance (millions)\end{tabular} & \multicolumn{4}{c}{9.0389} & 9.0389 \\
\begin{tabular}[c]{@{}l@{}}Buyer's risk \\ reduction (millions)\end{tabular} & \multicolumn{4}{c}{4.6595} & 4.6595 \\ \hline
\begin{tabular}[c]{@{}l@{}}Seller's risk without \\ insurance (millions)\end{tabular} & \multicolumn{4}{c}{0.0000} & 0.0000 \\
\begin{tabular}[c]{@{}l@{}}Seller's risk with \\ insurance (millions)\end{tabular} & \multicolumn{4}{c}{2.2977} & 2.2992 \\
\begin{tabular}[c]{@{}l@{}}Seller's risk \\ increase (millions)\end{tabular} & \multicolumn{4}{c}{2.2977} & 2.2992 \\ \hline
\begin{tabular}[c]{@{}l@{}}Aggregate risk without \\ insurance (millions)\end{tabular} & \multicolumn{4}{c}{13.6984} & 13.6984 \\
\begin{tabular}[c]{@{}l@{}}Aggregate risk with \\ insurance (millions)\end{tabular} & \multicolumn{4}{c}{11.3366} & 11.3381 \\
\begin{tabular}[c]{@{}l@{}}Aggregate risk \\ reduction (millions)\end{tabular} & \multicolumn{4}{c}{2.3618} & 2.3603 \\ \hline
\begin{tabular}[c]{@{}l@{}}Premium range \\ (millions)\end{tabular} & \multicolumn{4}{c}{{[}2.2977, 4.6595{]}} & {[}2.2992, 4.6595{]} \\ \hline
\end{tabular}%
}
\caption{Results of five trials of CEM with the same set of predicted incident probabilities.}
\label{tab:optim_results}
\end{table}

\section{Function Approximation} \label{sec:fun_approx}

In Section \ref{sec:implement}, two computational challenges are discussed in the implementation of the optimization algorithm, including the numerical evaluation of quantiles of loss random variables and random states that lead to non-convergent trials. Both challenges induce long running time and could make this policy design less practical in the production environment. To speed up the computation process, we make use of function approximation to find a \textit{target function}, which establishes a more direct mapping between incident properties, including their occurrence probabilities and severities, and the parameters of the eventual indemnity functions in the incident-specific policy. Note that in this study, the distribution parameters of incident severities are fixed, and only probabilities vary depending on company characteristics; therefore, more precisely speaking, the mapping is only between incident probabilities and indemnity function parameters. 

In this section, we distinguish between \textit{true solutions}, which are obtained by using CEM to approximately solve Problem \eqref{eq:varobj}, and \textit{fitted solutions}, which are obtained through function approximation. By using fitted solutions to approximate true solutions, such that the fitted solutions result in the same optimum in Problem \eqref{eq:varobj} as the true solutions, the end goal is that once a company's characteristics are fed into the trained stacking classifier built in Section \ref{sec:classification}, the corresponding probability vector can be used as inputs of the approximate function to generate the parameters of the optimal incident-specific cyber insurance, \textit{i.e.}, $\hat{\bm{\theta}}$ and $\hat{\bm{d}}$.

\subsection{Sample selection} \label{sec:fa_sample}
Provided that the true solutions to Problem \eqref{eq:varobj} are non-unique, as shown in Section \ref{sec:cme}, the mapping between $\bm{p}$ and $\left(\bm{\theta}^*,\bm{d}^*\right)$ is one-to-many. This could potentially lead to a lack of fit and suboptimal fitted solutions. The suboptimality is a result of the fact that the true solutions of individual $d^*_k$, for $k = 1,\dots, K$, are on disjoint intervals on the real line (see \cite{Asimit2021}) but the function approximation process can easily disregard that premise if, with respect to the same set of probabilities, the divergence between the fitted solution and all true solutions, \textit{e.g.}, mean squared error or mean absolute error, is to be minimized because the fitted value could plausibly be outside the intervals where the true solutions reside in. 

Therefore, instead of looking for a fitted solution with the smallest average deviation from all true solutions, it is more guaranteed to find an optimal fitted solution close to one of the true solutions. In this study, to identify a unique true solution, we choose the one that minimizes the seller's expected loss among non-unique true solutions. That is, the solution selected for function approximation is, therefore, ${\arg \min}_{(\bm{d}, \bm{\theta}) \in \mathcal{S}} \mathbb{E}L_s(\bm{d}, \bm{\theta}; X, O) $, where $\mathcal{S}$ is the set of Pareto optimal $\left(\bm{\theta}^*,\bm{d}^*\right)$ computed by the CEM. 
%$\hat{\mathcal{S}}$ be the set of all computed true solutions with respect to a certain probability vector $\bm{p}$, which is a subset of all true solutions $\mathcal{S} = {\arg \min}_{\substack{\bm{d} \in \mathbb{R}_+^K, \\ \bm{\theta} \in \{0,1\}^K}} F(\bm{d}, \bm{\theta}; X, O)$, and .
This strategy serves as an illustration of how a unique true solution can be chosen based on additional preferences, while in practice, alternative preferences can be used to compare among the non-unique solutions to Problem \eqref{eq:varobj} and choose the best one. 

\subsection{Function approximation procedure}
Because $\bm{d}$ depends on $\bm{\theta}$, the function approximation follows a two-step process. The relationship between probabilities $\bm{p}$ and the types of indemnities $\bm{\theta}$ is first approximated, for which $\bm{p}$ are seen as features, and $\bm{\theta}$ are seen as labels. Because each $\theta_k$, for $k = 1,\dots, K$, takes the value of either $0$ or $1$ and is not mutually exclusive, predicting $\bm{\theta}$ is a multi-label classification problem. Then, conditioning on $\bm{\theta}$, the mapping between probabilities $\bm{p}$ and the amounts of incident-specific deductibles or limits $\bm{d}$ shall be approximated, and this is a multi-output regression problem. Because there are $2^K$ distinct values of $\bm{\theta}$, the number of regression models to build is $2^K$. The target function to be approximated then consists of both the classification model and the conditional regression models. 

Let $G \colon (0,1)^K \rightarrow \{0,1\}^K$ be the classifier and $H_{\bm{\theta}} \colon (0,1)^K \rightarrow \mathbb{R}_+^K$ be the conditional regression models depending on $\bm{\theta}$. In addition, the sample created in Section \ref{sec:fa_sample} is split into a training set $\{(\bm{p}_s;\bm{\theta}^*_s, \bm{d}^*_s)\}_{s=1}^{S_1}$ and a test set $\{(\bm{p}_{S_1+s};\bm{\theta}^*_{S_1+s}, \bm{d}^*_{S_1+s})\}_{s=1}^{S_2}$, where $S_1$ and $S_2$ are the sizes of the two sets, respectively. The function approximation procedure can be summarized in Algorithm \ref{alg:fa}.

\begin{algorithm}[!ht]
\SetNoFillComment
\caption{Function Approximation: Model training and testing}\label{alg:fa}
\Comment{Training}
\KwIn{$\{(\bm{p}_s;\bm{\theta}^*_s, \bm{d}^*_s)\}_{s=1}^{S_1}$, initial classifier $G^0$, initial regression models $H^0_{\bm{\theta}}$}

\KwOut{Trained classifier $G^*$, trained regression models $H^*_{\bm{\theta}}$}

Training the multi-label classifier $G^0$ on the training set with $\bm{p}$ as features and $\bm{\theta}$ as labels

\For{ $\bm{\theta} \in \{0,1\}^K$}{
Training the multi-output regression model $H^0_{\bm{\theta}}$ on the training set with $\bm{p}$ as features and $\bm{d}$ as labels
}

\Comment{Testing}
\KwIn{$\{(\bm{p}_{S_1+s}; \bm{\theta}_{S_1+s}^*, \bm{d}_{S_1+s}^*)\}_{s=1}^{S_2}$, trained classifier $G^*$, trained regression models $H^*_{\bm{\theta}}$}

\KwOut{Error rate, fitted solutions $\{(\hat{\bm{\theta}}_{S_1+s}, \hat{\bm{d}}_{S_1+s})\}_{s=1}^{S_2}$}

\For{$s = 1,2,\dots,S_2$}{
    $\hat{\bm{\theta}}_{S_1+s} \gets G^*(\bm{p}_{S_1+s})$
    
    $\hat{\bm{d}}_{S_1+s} \gets H_{\hat{\bm{\theta}}_{S_1+s}}^*(\bm{p}_{S_1+s})$
    
    $\epsilon_{S_1+s} \gets F(\bm{d}^*_{S_1+s}, \bm{\theta}^*_{S_1+s}; X, O) - F(\hat{\bm{d}}_{S_1+s}, \hat{\bm{\theta}}_{S_1+s}; X, O)$ 
}

Error rate $\gets \left(\sum_{s = 1}^{S_2} \mathbb{I}_{\{\epsilon_{S_1+s} \ne 0\}}\right)\big/{S_2}$
\end{algorithm}

For the evaluation of the fitted solutions produced on the test set, error is defined as the difference between the values of the objective function in Problem \eqref{eq:varobj} when true solutions and fitted solutions are provided, respectively, \textit{i.e.}, ${\epsilon_s = F(\bm{d}^*_s, \bm{\theta}^*_s; X, O) - F(\hat{\bm{d}}_s, \hat{\bm{\theta}}_s; X, O)}$. The error rate is used as a performance measure of the target function, which represents the ratio of predictions made on the test set resulting in the same optimums as true solutions. 

\subsection{Numerical results}
For each distinct set of probabilities $\bm{p}$, $50$ corresponding true solutions are computed using the CEM, each of which has a different random state, and then among them, a unique true solution, which minimizes the seller's expected loss, is selected for the sample in function approximation. 

The computation of individual optimization tasks with specifications as shown in Appendix \ref{append:params} is parallelized on $96$ CPU cores. During each 24-hour wall-clock time, the average number of unique true solutions that can be generated is around $790$. This computationally intense process highlights the necessity of the use of function approximation in practice. 
For both the classification and regression tasks, we build tree-based models using the LightGBM framework (see \citet{ke2017lightgbm}), with a training set of size 2,201. Table \ref{tab:fa_num_results} shows the fitted solutions of five organizations, provided their organizational characteristics. 

\begin{table}[]
\centering
\resizebox{\textwidth}{!}{%
\begin{tabular}{lrrrrr}
\hline
\textbf{Organization} & \textbf{1} & \textbf{2} & \textbf{3} & \textbf{4} & \textbf{5} \\ \hline
NCASE & 55 & 652 & 28 & 3300 & 270 \\
EMP & 4500 & 25105 & 119 & 16000 & 7908 \\
REV (millions) & 461.47 & 40653.00 & 0.72 & 2662.68 & 1448.391 \\
CTYPE & Private & Public & Private & Public & Other \\
STATE & Other & California & Arizona & Connecticut & Florida \\
IND & Services & Services & Retail Trade & \begin{tabular}[c]{@{}r@{}}Transportation \& \\ Public Utilities\end{tabular} & \begin{tabular}[c]{@{}r@{}}Public \\ Administration\end{tabular} \\
\multicolumn{6}{c}{\textbf{Incident Occurrence}} \\
AFTER2012 & Yes & Yes & Yes & Yes & Yes \\
MON & January & January & January & March & March \\ \hline
$p_1$ & 0.3383 & 0.4401 & 0.4700 & 0.4340 & 0.2300 \\
$p_2$ & 0.5717 & 0.3340 & 0.3400 & 0.4360 & 0.4800 \\
$p_3$ & 0.0700 & 0.1764 & 0.1600 & 0.0600 & 0.1900 \\
$p_4$ & 0.0200 & 0.0495 & 0.0300 & 0.0700 & 0.1000 \\ \hline
$\hat{\theta}_1$ & 0 & 1 & 1 & 0 & 0 \\
$\hat{\theta}_2$ & 1 & 1 & 1 & 1 & 1 \\
$\hat{\theta}_3$ & 1 & 0 & 0 & 1 & 1 \\
$\hat{\theta}_4$ & 1 & 0 & 0 & 1 & 0 \\ \hline
$\hat{d}_1$ (millions) & 0.1430 & 0.0968 & 0.0943 & 0.1318 & 0.1094 \\
$\hat{d}_2$ (millions) & 0.1502 & 0.0901 & 0.0924 & 0.1636 & 0.1852 \\
$\hat{d}_3$ (millions) & 0.0957 & 0.0904 & 0.0781 & 0.0918 & 0.1477 \\
$\hat{d}_4$ (millions) & 0.0768 & 0.1490 & 0.1249 & 0.0974 & 0.1228 \\ \hline
\begin{tabular}[c]{@{}l@{}}CEM optimum \\ \;(millions)\end{tabular} & 11.3366 & 7.4182 & 7.7513 & 9.7930 & 9.2889 \\
\begin{tabular}[c]{@{}l@{}}Function \\ \;approximation \\ \;optimum (millions)\end{tabular} & 11.3366 & 7.4197 & 7.7513 & 9.7930 & 9.2889 \\
\begin{tabular}[c]{@{}l@{}}Error \\ \;(millions)\end{tabular} & 0.0000 & 0.0015 & 0.0000 & 0.0000 & 0.0000 \\ \hline
\begin{tabular}[c]{@{}l@{}}Buyer's risk without \\ \;insurance (millions)\end{tabular} & 13.6984 & 8.5989 & 8.7049 & 11.4981 & 11.3691 \\
\begin{tabular}[c]{@{}l@{}}Buyer's risk with \\ \;insurance (millions)\end{tabular} & 9.0389 & 6.9697 & 7.4837 & 5.9956 & 6.5337 \\
\begin{tabular}[c]{@{}l@{}}Buyer's risk \\ reduction (millions)\end{tabular} & 4.6595 & 1.6292 & 1.2212 & 5.5025 & 4.8354 \\ \hline
\begin{tabular}[c]{@{}l@{}}Seller's risk without \\ \;insurance (millions)\end{tabular} & 0.0000 & 0.0000 & 0.0000 & 0.0000 & 0.0000 \\
\begin{tabular}[c]{@{}l@{}}Seller's risk with \\ \;insurance (millions)\end{tabular} & 2.2977 & 0.4500 & 0.2676 & 3.7974 & 2.7552 \\
\begin{tabular}[c]{@{}l@{}}Seller's risk \\ \;increase (millions)\end{tabular} & 2.2977 & 0.4500 & 0.2676 & 3.7974 & 2.7552 \\ \hline
\begin{tabular}[c]{@{}l@{}}Aggregate risk without \\ \;insurance (millions)\end{tabular} & 13.6984 & 8.5989 & 8.7049 & 11.4981 & 11.3691 \\
\begin{tabular}[c]{@{}l@{}}Aggregate risk with \\ \;insurance (millions)\end{tabular} & 11.3366 & 7.4197 & 7.7513 & 9.7930 & 9.2889 \\
\begin{tabular}[c]{@{}l@{}}Aggregate risk \\ \;reduction (millions)\end{tabular} & 2.3618 & 1.1792 & 0.9536 & 1.7051 & 2.0802 \\ \hline
\begin{tabular}[c]{@{}l@{}}Premium range \\ \;(millions)\end{tabular} & {[}2.2977, 4.6595{]} & {[}0.4500, 1.6292{]} & {[}0.2676, 1.2212{]} & {[}3.7974, 5.5025{]} & {[}2.7552, 4.8354{]} \\ \hline
\end{tabular}%
}
\caption{Comparisons between exact solutions solved by the CEM and fitted solutions by function approximation.}
\label{tab:fa_num_results}
\end{table}

This table summarizes multiple aspects of the workflow proposed by this paper. First, there are indeed different optimal incident-specific insurance designs for organizations of distinct characteristics. The presented organizations vary in litigation history, size, ownership, location, and industry. These characteristics can be used as rating factors in the underwriting process, and the insurer should be able to collect these pieces of information on the insurance buyer conveniently. This result validates the practice of coverage being designed in an incident-specific manner.

Second, similar to the results shown in Table \ref{tab:optim_results} of Subsection \ref{sec:implement}, Table \ref{tab:fa_num_results} demonstrates how both the insured and the insurer can be mutually benefited from an incident-specific cyber policy. Indeed, Organization 1 is the same organization used to derive the results in Table \ref{tab:optim_results}, and the numbers regarding risks with and without insurance and the premium range are identical in both tables. We shall relegate readers to Subsection \ref{sec:implement} for a detailed account of the policy being mutually beneficial to both parties. In addition, the same conclusion can be drawn from the other four organizations presented in Table \ref{tab:fa_num_results}, suggesting that the proposed incident-specific policy can offer such a benefit to organizations of different characteristics. However, Table \ref{tab:fa_num_results} shows that the effectiveness of aggregate cyber risk reduction does not always hold. Indeed, while the premiums to be charged for Organizations 1, 2, and 3 could be lower than their respective aggregate cyber risk reductions, the premiums to be charged for Organizations 4 and 5 must be larger than their respective aggregate risk reductions.

% Table \ref{tab:fa_num_results} demonstrates that the proposed risk-sharing scheme reduces the overall risk borne by the policyholder and the insurer. For example, when Organization 1 purchases no insurance, its risk, measured according to the chosen risk measure, is \$13.6984 million. After it and an insurer enter into an incident-specific cyber insurance contract designed by following the workflow described in previous sections, the risk taken by both parties is reduced by \$2.3618 million to \$11.3366 million. In this situation, with an appropriate amount of premium being paid by the policyholder according to Inequality \eqref{eq:premium_range}, both parties can be better off than the case without insurance. This risk reduction is also found in other organizations if they are covered by the proposed optimal incident-specific insurance.

Lastly, in most cases, the target function can produce solutions that yield the exact optimum as the true solutions generated by the CEM. A small percentage of the fitted solutions are non-optimal; see, for example, Organization 2. This problem can potentially be mitigated by employing a larger training set to train the target function. Overall, an error rate of $0.22$ on the test set of size 790 is attained. That is, 78\% of the fitted solutions result in the same optimal risks as those given by true solutions. The slight compromise in accuracy leads to high computational efficiency. For a test set of size $5$, the time needed for computing the fitted solutions is approximately $0.068$ seconds. The running time here and the running time of the CEM presented in Table \ref{tab:time_cem} are measured on the same hardware. Compared to the aforementioned CEM, which may take hours to solve the problem, the usage of function approximation makes it much more feasible to generate policy specifications on the fly in the production environment. 

\section{Conclusions and Future Directions}
In this paper, we proposed a workflow for the design of incident-specific cyber insurance. It consists of three key components, including the estimation of model inputs using public and proprietary data in the underwriting process, the Pareto optimal objective based on the insured's and the insurer's preference orderings by their risk measures, and solvers for the optimization problem. Using real cyber incident data, we showed how this workflow generates incident-specific policies that lower the total risk taken by the insured and the insurer. We also demonstrated that the proposed workflow can be time efficient with the help of function approximation if there is a time constraint on the delivery of results, which is common for insurers that provide quotes and coverages on the fly. 

Despite the time efficiency of using an approximated target function, one limitation is that the optimum is not always attained by the fitted solutions generated by the target function. Therefore, the benefits and risks of using this approach should be weighed in practice. Another limitation of this study is that, because of the scarcity of data on cyber incident losses, it is not feasible to model company-specific loss distributions for individual incident types, and therefore we only considered how incident type realizations are related to company characteristics, regardless of the relationship between those characteristics and the potential cyber losses. This issue could possibly be resolved in future studies if a better quality of cyber loss dataset is available. Lastly, this study assumes that indemnity functions are specific to cyber incident types, or equivalently perils, only. In practice, deductibles and limits could also be placed on more granular levels like covered assets, such as hardware, data, or business income; see, for example, \citet{chong2022cyber}. Depending on the use case, the proposed workflow can easily be adopted for policies that offer coverages at such more granular level.

\section*{Acknowledgments}
Wing Fung Chong, Dani{\"e}l Linders, and Linfeng Zhang are supported by a 2018 Ignacio Hernando de Larramendi Research Grant from the Fundaci{\'o}n MAPFRE. Wing Fung Chong, Dani{\"e}l Linders, Zhiyu Quan, and Linfeng Zhang are supported by a Centers of Actuarial Excellence (CAE) Research Grant (2019–2021) from the Society of Actuaries (SOA). Any opinions, findings, conclusions, or recommendations expressed in this material are those of the authors and do not necessarily reflect the views of the Fundaci{\'o}n MAPFRE and the SOA.

\newpage

\printbibliography

\newpage

\begin{appendices}
\section{Explanatory Variables and Their Descriptions}\label{append_explan}
\begin{table}[!h]
\centering
\begin{tabular}{@{}ll@{}}
\toprule
\textbf{Explanatory Variable} & \multicolumn{1}{c}{\textbf{Description}}                \\ \midrule
EMP               & Number of employees                                     \\
NCASE             & Number of Federal Docket cases linked to company        \\
REV               & Total revenue (in millions of USD)                      \\
MON               & Month in which the incident occurs                      \\
STATE             & State in which the victim company is based              \\
CTYPE             & Company type                                            \\
                  & * PRV (Private)                                         \\
                  & * PUB (Public)                                          \\
IND               & Industy by SIC divisions                                \\
YEAR              & Whether the incident occurred before 2012 or after 2012 \\ \bottomrule
\end{tabular}
\caption{All explanatory variables and their descriptions.}
\label{tab:var_desc}
\end{table}
\newpage

\section{Classification Models}\label{append:classifiers}
\subparagraph{Decision tree}
A decision tree (DT) splits the dataset based on the values of explanatory variables and attempts to partition the dataset into homogeneous subsets. Each split of a parent node is performed with the goal of reducing the Gini impurity of data in child nodes. Decision tree is known to have a tendency to overfit, which means that it can fit the training data well by growing a complex tree with granular partitions, but its performance can suffer when making predictions on new data. To mitigate this issue, three hyperparameters, including the maximum depth of the tree, the minimum number of samples required in each leaf, and the complexity parameter of the tree pruning process, are tuned by grid search with 5-fold cross validations, to improve the out-of-sample prediction performance of the built tree. The hyperparameter tuning process of all other classification methods mentioned in this paper are performed in the same way.

\subparagraph{Random forest}
Random Forest (RF) is an ensemble model that employs multiple classification trees, of which each is built based on a bootstrapped sample of the training set and a randomly selected subset of all explanatory variables. In the end, the probabilistic predictions made by all trees are averaged to give a final prediction. The randomness introduced by bootstrapping and randomly selected explanatory variables help reduce the variance of the model, thus mitigating the overfitting issue of a single classification tree, as discussed previously. Hyperparameters including the number of trees to build and the size of the subset of explanatory variables are tuned for performance improvement in this study. 

\subparagraph{Gradient boosted trees}
Gradient boosted trees is another type of ensemble model based on decision trees. Instead of growing trees independently and then aggregating the results, the boosted method iteratively improves the performance of the previous tree-based classifier by updating it with a new tree that fits the residual. The updating process is a linear combination of the old classifier and the newly built tree, in which the coefficients minimize a certain loss function of the true values and the predicted values. Hyperparameters, including the learning rate, which governs how much each new tree contribute to the final classifier, and the total number of trees to build, are tuned.

\subparagraph{Linear discriminant analysis}
Linear discriminant analysis (LDA) solves an optimization problem, for which a transformation matrix is solved such that after the transformation, the ratio between the total between-class variance of data points measured in an Euclidean space and the total within-class variance is maximized. That is, data points in the same class are made as close to each other as possible, whereas data points belonging to different classes are separated to the largest extent. The advantage of this method over the three tree-based models introduced earlier is that there is an analytical solution to the transformation matrix, and thus the training time is minimal. The drawback, however, is that it imposes relatively strict assumptions on the data, such as multivariate normality, independence, and non-collinearity, which can rarely be fully satisfied by empirical data, and in that case, the performance can suffer. Given that our sample contains many dummy variables, the assumptions are largely violated. Nevertheless, this method is tested and compared with other models. 

\subparagraph{Multinomial logistic regression}
For multi-class problems, multinomial logistic regression (MLR) is a generalization of its binary counterpart. Let $K$ denote the number of classes and for each class $k=1,2,\dots,K$, the probability $p_k$ of an observation with features vector $\bm{x}$ belonging to that class is modeled as follows,
$$
\ln p_k = \beta_k \cdot \bm{x} - \ln \sum_{k=1}^K e^{\beta_k \cdot \bm{x}},
$$
where $\beta_k$ is the coefficient learned from the training set to minimize the cross-entropy loss function, and the term $\ln \sum_{k=1}^K e^{\beta_k \cdot \bm{x}}$ is a normalizing factor that ensures $\sum_{k=1}^K p_k=1$. Equivalently, 
$$
p_k = \frac{e^{\beta_k\cdot\bm{x}}}{\sum_{k=1}^K e^{\beta_k \cdot \bm{x}}},\; \text{ for } k = 1,2,\dots,K
$$
which is the softmax function of $\beta_k \cdot \bm{x}$.

\subparagraph{Multi-layer perceptron}
Multinomial logistic regression can be viewed as a single-layer perceptron without hidden layers, of which the activation function in the output layer is softmax, as introduced earlier. Thus, it would be interesting to compare multinomial logistic regression to multi-layer perceptron (MLP) models with additional hidden layers. Hidden layers enable the model to learn non-linear relationships, and thus might provide a better fitting. For experiment purposes, an MLP classifier with two hidden layers and ten neurons in each hidden layer is trained.

\subparagraph{Stacking classifier}
Lastly, stacking classifiers are built based on the results output by the different algorithms mentioned above. The motivation is that classifiers of different kinds are likely to produce errors in different ways, and thus combining their results with proper weights may result in reduced errors, as suggested in \citet{Merz1999}. To stack the base classifiers as introduced previously, their probabilistic predictions are collected and used as new features for building meta-classifiers on top of all individual models. The choice of the meta-classifier in this study is random forest. To see how different sets of base classifiers can affect the performance of the meta-classifier, three stacking classifiers are compared. The first uses only tree-based base classifiers, including decision trees, random forests, and gradient-boosted trees. The second stacking classifier excludes the two linear methods, which are linear discriminant analysis and multinomial logistic regression. The third stacking classifier includes all models mentioned previously as base classifiers.

\newpage

\section{Optimization Parameters} \label{append:params} 
\begin{table}[htbp]
\centering
\begin{tabular}%
{>{\raggedright\arraybackslash}p{0.4\linewidth}%
>{\raggedleft\arraybackslash}p{0.4\linewidth}%
}
\toprule
\textbf{Model Inputs} & \textbf{Value} \\ \midrule
Seller's Risk Level $\alpha$ & 0.95 \\
Buyer's Risk Level $\beta$ & 0.9 \\ \midrule
\textbf{CEM Specifications} & \textbf{Value} \\ \midrule
Initial Distribution of $\bm{d}$ & Truncated multivariate normal with the mean and standard deviation of each marginal distribution to be 0 and 100,000 respectively \\
Initial Distribution of $\bm{\theta}$ & Multivariate Bernoulli with the success probability of each marginal distribution to be 0.5 \\
Variance threshold of $\bm{d}$ & 0.1 \\
Variance threshold of $\bm{\theta}$ & 0.01 \\
Sample size $N$ & 10 \\
Elite sample proportion $\varrho$ & 0.2 \\
No-improve iterations before stop $l$ & 10 \\
Maximum Number of iterations & 401 \\ \bottomrule
\end{tabular}
\caption{Seller's and buyer's risk levels and CEM specifications}
\label{tab:algo_spec}
\end{table}

\newpage

\section{Severity Distribution Fitting Results}\label{append_gof}
\subsection{Comparison of AICs of fitted distributions}\label{append_aics}
% \usepackage{booktabs}
% \usepackage{booktabs}
\begin{table}[h]
\begin{tabular}{@{}l|rrrr}
\toprule
                     & \multicolumn{1}{c}{\textbf{Privacy Violation}} & \multicolumn{1}{c}{\textbf{Data Breach}} & \multicolumn{1}{c}{\textbf{Extortion/Fraud}} & \multicolumn{1}{c}{\textbf{IT Error}} \\ \midrule
\textit{log-normal}  & \textbf{32.82}                                 & \textbf{2154.44}                         & \textbf{-2135.91}                            & \textbf{294.72}                       \\
\textit{exponential} & 9422.81                                        & 5014.21                                  & 4259.55                                      & 1714.78                               \\
\textit{gamma}       & 939.88                                         & 2484.55                                  & -1167.95                                     & 454.25                                \\
\textit{weibull}     & 269.59                                         & 2228.09                                  & -1896.31                                     & 352.69                                \\ \bottomrule
\end{tabular}
\caption{AICs of distributions fitted to incident-specific losses.  For each incident type, the log-normal distribution has the lowest value among all fitted distributions.}
\label{tab:fitted_aics}
\end{table}
% \subsection{Goodness-of-fit of fitted log-normal distributions}\label{append_gof_plots}

% \begin{figure}[htbp]
%     \centering
%     \includegraphics[scale = 0.9]{fit4_gof.pdf}
%     \caption{Fitted loss distribution of privacy violation (PV) incidents.}
%     \label{fig:fit4_gof}
% \end{figure}
% \begin{figure}[htbp]
%     \centering
%     \includegraphics[scale = 0.9]{fit1_gof.pdf}
%     \caption{Fitted loss distribution of data breach (DB) incidents.}
%     \label{fig:fit1_gof}
% \end{figure}
% \begin{figure}[htbp]
%     \centering
%     \includegraphics[scale = 0.9]{fit2_gof.pdf}
%     \caption{Fitted loss distribution of fraud and extortion (FE) incidents.}
%     \label{fig:fit2_gof}
% \end{figure}
% \begin{figure}[htbp]
%     \centering
%     \includegraphics[scale = 0.9]{fit3_gof.pdf}
%     \caption{Fitted loss distribution of IT error (ITE) incidents.}
%     \label{fig:fit3_gof}
% \end{figure}

\end{appendices}

\end{document}